\documentclass{article}
\usepackage[latin1]{inputenc}
\usepackage{amsmath}
\usepackage{graphicx}

\newcommand{\mean}[1]{\left< #1 \right>}
\newcommand{\graphWidth}{4.35cm}
\newcommand{\eqIndent}{3.5cm}
\newcommand{\rn}{{\textrm{r}}}

\newcommand{\ncnI}{{\textrm{nc},1}}
\newcommand{\ncnII}{{\textrm{nc},2}}
\newcommand{\cn}{{\textrm{c}}}
\newcommand{\Pn}{{P_{\textrm{n}}}}
\newcommand{\Pm}{{P_{\textrm{m}}}}
\newcommand{\Pt}{{P_{\textrm{t}}}}
\newcommand{\GCC}[1]{{\textrm{GCC}_{#1}}}
\newcommand{\ZGCC}[2]{{Z_\GCC{#2}^{(#1)}}}
\newcommand{\Beta}{{\mathrm{B}}}

\begin{document}

\title{Local Heuristics and the Emergence of Spanning Subgraphs in Complex Networks}

\author{Alexandre~O.~Stauffer\\
Valmir~C.~Barbosa\thanks{Corresponding author ({\tt valmir@cos.ufrj.br}).}\\
\\
Universidade Federal do Rio de Janeiro\\
Programa de Engenharia de Sistemas e Computa\c c\~ao, COPPE\\
Caixa Postal 68511\\
21941-972 Rio de Janeiro - RJ, Brazil}

\date{November 24, 2004}

\maketitle

\begin{abstract}
We study the use of local heuristics to determine spanning subgraphs for use in
the dissemination
of information in complex networks. We introduce two different heuristics and analyze their behavior
in giving rise to spanning subgraphs that perform well in terms of allowing every node of the network
to be reached, of requiring relatively few messages and small node bandwidth for information
dissemination, and also of stretching paths with respect to the underlying network only modestly.
We contribute a detailed mathematical analysis of one of the heuristics and provide extensive simulation
results on random graphs for both of them. These results indicate that, within certain limits, spanning subgraphs are indeed
expected to emerge that perform well in respect to all requirements. We also discuss the spanning subgraphs'
inherent resilience to failures and adaptability to topological changes.

\end{abstract}

\section{Introduction} \label{sec:introduction}
Let $G=(N_G,E_G)$ be an undirected graph with $n=\vert N_G\vert$ nodes
and edges representing bidirectional
links for pairwise communication among the nodes. We regard $G$ as standing for some unstructured,
real-world network whose nodes have no more information on the overall topology
of $G$ than can be inferred from their immediate neighborhoods. Given these
characteristics, $G$ can also be seen as belonging to the class of networks that
have recently come to be referred to as complex networks \cite{albert2002}.

Several of the typical problems that require a distributed solution by the nodes
of $G$ frequently involve the need to disseminate a piece of information, call
it $I$, through the nodes of the network that share the same connected component
with the node that originally possesses $I$. We assume that this node is unique
with respect to that particular information dissemination and refer to it as the
originator.

There is a host of possibilities to solve this basic problem of disseminating
information through the nodes of $G$, but invariably they either depend on the
existence of a spanning subgraph of $G$ on whose edges the dissemination is
performed, or else they employ straightforward flooding of the network's edges
by copies of $I$. The former alternative is often regarded as substantially
more cost-effective in terms of several quantities of interest, but of course
it carries with it the inherent need for the desired spanning subgraph to be
initially determined and subsequently maintained if the network undergoes
topological changes \cite{barbosa1996}.

Subgraphs of interest in this context include the well-known minimum spanning
trees, for which several procedures related to creation and maintenance are
available \cite{awerbuch1987,cheng1988,faloutsos1995}, and include also the more general, so-called spanners, which bring
with them well-defined structural requirements related to efficiency indicators,
but are on the other hand considerably less well-known \cite{kortsarz1998,peleg2000}. But regardless of the
particular guise of the spanning subgraphs of $G$ for use in information
dissemination, the importance of studying them in detail has in recent years
found strong justification from the practical side. Notable examples here
include the case in which $G$ is some virtual supergraph of a physical network;
in this case, spanning subgraphs of $G$ are needed to function as the so-called
overlay networks for end-to-end communication over the underlying physical
network \cite{chu2000}.

In this paper we focus on one very basic question related to determining a
spanning subgraph of a complex network $G$: how close can we get to obtaining
a subgraph of $G$ that can be used to disseminate $I$ through the nodes of the
originator's connected component, while at the same time satisfying some basic
set of performance requirements, if nodes are only allowed to use local
information (i.e., information that can be obtained from no farther than the
nodes' immediate neighborhoods)? Important requirements involve the expected
number of nodes reached when $I$ is disseminated, the expected number of
copies of $I$ that are needed, the expected degree of each node in the subgraph (since
it relates closely to how many copies of $I$ a node can concurrently send out
given its bandwidth limitations), and also the expected path length from the originator on
the subgraph.

Even though of a fundamental nature, this question is admittedly too general for
an objective analysis. For this reason, we concentrate on the narrower issue of
investigating what happens in terms of the aforementioned performance
requirements when a subgraph of $G$ is determined in a fully distributed fashion
by the nodes according to the following strictly local rules. Each node is
responsible for choosing some of its own neighbors in the subgraph and makes its
choices in two subsequent steps: first one node is picked from among the node's
neighbors; then, with some nonzero probability that we call $\alpha$, the node
gets to pick a second neighbor (which we allow to be identical to the first one
it picked).

If this simple procedure is performed by all the nodes of $G$, then clearly the
resulting subgraph, which we denote by $D=(N_D,E_D)$, has $N_D=N_G$ while $E_D$
contains all the edges $(u,v)$ from $G$ such that $u$ chose $v$ at least once
or $v$ chose $u$. This subgraph is then necessarily a spanning subgraph of $G$;
using it for disseminating $I$ from the originator is simply a question of
having the originator send $I$ to all of its neighbors in $D$, and similarly for
all the other nodes when they receive $I$ for the first time. Our four
performance requirements now have to be examined in terms of how the connected
components of $D$ relate to those of $G$. In particular, does the connected
component of $D$ to which the originator belongs span the entire connected
component of $G$ that contains it?

Throughout the paper we use $c_u$ to denote the number of choices made by node
$u$ and $C_u$ to denote the set of nodes (some of $u$'s neighbors) that get
chosen by $u$. Clearly, $1\le\vert C_u\vert\le c_u\le 2$ and the set of $u$'s
neighbors in $D$ is given by $C_u$, possibly enlarged by every other node $v$
that is a neighbor of $u$ in $G$ and such that $u\in C_v$.

We call $D$ a dissemination subgraph of $G$ and devote the remainder of the
paper to analyzing its properties. Our analysis depends, naturally, on the
specific criteria that each node uses when making its two decisions. We consider
two possibilities, of which the simplest, referred to as the uniform approach,
lets each node make its choices uniformly at random among its neighbors. Our
analytical treatment of $D$'s properties in this case is given in
Section~\ref{sec:analysis}; it is based on regarding $G$ as a random graph and makes use of
the principles laid down in \cite{newman2001}. This is our core section, and its results are
complemented by the simulation results we present in Section~\ref{sec:simulation} for
Poisson-distributed node degrees (the classic Erd\H{o}s-R\'{e}nyi model \cite{erdos1959}) and
also for degrees distributed according to a power law (recently discovered to be
approximately representative of relevant real-world networks \cite{faloutsos1999,medina2000,albert2002}).

Even though the analytical treatment we offer in Section~\ref{sec:analysis} is specific to
this simplest possibility for a choice criterion, at this point it seems to be
as far as we have the means to go. For this reason, and notwithstanding the
second possibility's clear superiority in terms of our stated performance requirements (see below),
we only treat that possibility by means of simulations, whose results are described in
Section~\ref{sec:simulation} along with those for the uniform approach. In any event, our
mathematical analysis for the uniform approach is innovative and we believe it
may yield interesting insight into the analysis of similar problems on complex
networks. It may also ultimately be possible to generalize it to handle the
more complicated case of the second possibility of choice criteria.

We refer to the second possibility as the degree-based approach. In this
approach, the first choice by a node selects uniformly at random from those of
its neighbors that have the highest degree (if this is the case for only one
neighbor, then the first choice degenerates into a deterministic decision). The
second choice selects a neighbor randomly in proportion to its degree. The
degree-based approach is clearly much less uninformed on the network's topology
than the uniform approach. For this reason, it is expected to surpass the
uniform approach in terms of the indicators we have informally introduced. That
this is indeed the case is apparent from the simulation results we show in
Section~\ref{sec:simulation}. But, even if we find this to be only expected, we
also find it remarkable that such a simple strategy of
strictly local nature should support a positive answer to our original
question to the extent that it does.

We complement our study of our local, two-choice scheme to approximate a
spanning subgraph of $G$ by elaborating on its resilience and adaptability
properties. These are crucial in the context of networks that may undergo
topological changes and we treat them in Section~\ref{sec:dynamicTopology}. To finalize, we
offer concluding remarks in Section~\ref{sec:conclusions}.

\section{Mathematical analysis} \label{sec:analysis}

Henceforth we regard $G$ as a random graph having node degrees distributed independently from one another and identically to a random
variable $K_G$. Furthermore, the nodes of $G$ are assumed to be connected to one another at random given their degrees,
so the degrees of two adjacent nodes remain independent. Our results in this section target the case of a formally infinite set of
nodes, that is, the case in which $n \to \infty$.

Let $P_G(a)$ be the probability that a randomly chosen node of $G$ has degree $a$. The average degree of $G$, denoted by $Z_G$, is
$Z_G = \sum_{a=0}^{n-1} a P_G(a)$. Also, given the random nature of node interconnections in $G$, the probability that some node's
neighbor has degree $b$ is equal to the expected fraction of edges incident to degree-$b$
nodes, which is given by
\begin{equation}
   \frac{b P_G(b)}{\sum_{a=0}^{n-1} a P_G(a)} = \frac{b P_G(b)}{Z_G}.
   \label{eq:degneigh}
\end{equation}

From \cite{molloy1995,cohen2000,newman2001}, we know how to characterize the existence in $G$ of a large,
size-$\Theta(n)$ connected component, commonly known as the giant connected component of $G$ (henceforth denoted by $\GCC{G}$).
If $\mean{K_G^2}$ denotes the second moment of the random variable $K_G$,
that is, $\mean{K_G^2} = \sum_{a=0}^{n-1} a^2 P_G(a)$, then $\GCC{G}$ almost surely exists if and only if
\begin{equation}
   \frac{\mean{K_G^2}}{Z_G} > 2
   \label{eq:criterion}
\end{equation}
or, equivalently,
\begin{equation}
   \sum_{a=0}^{n-1} \frac{a P_G(a)}{Z_G} a > 2.
   \label{eq:criterion2}
\end{equation}
Intuitively, for a randomly chosen node $u$ and letting $v$ be one of its neighbors, this means that
$\GCC{G}$ almost surely exists (and then $G$ is said to be above the phase transition that gives rise to $\GCC{G}$) if
and only if $v$ is expected to have strictly more than one neighbor besides $u$. Otherwise, all the connected components of $G$ are small,
consisting of $o(n)$ nodes (and $G$ is said to be below the phase transition).

Now let $D$ be a dissemination subgraph of $G$ constructed by the uniform approach; it is also a random graph, and we let $\GCC{D}$ denote
its giant connected component. Given a randomly chosen, degree-$a$ node $u$ of $G$, let $\pi^{(a)}_\rn(1)$ denote the probability that
$\vert C_u \vert = 1$. If $a>0$, then
\begin{equation}
   \pi^{(a)}_\rn(1) = 1-\alpha + \alpha/a,
\end{equation}
which reflects the probability that either $c_u=1$ or $c_u=2$ but both of $u$'s choices were identical.
It follows that the probability that $\vert C_u \vert = 2$ is
\begin{equation}
   \pi^{(a)}_\rn(2) = 1 - \pi^{(a)}_\rn(1) = \alpha(a-1)/a.
\end{equation}
For $a=0$, clearly $\pi^{(a)}_\rn(1) = \pi^{(a)}_\rn(2) = 0$.

We now pause momentarily to note that, throughout the paper, we employ the mnemonic ``$\rn$'' when referring to randomly chosen nodes
as in the preceding paragraph. Furthermore, when considering a node $v$ reached by following
one of its incident edges, say $(u,v)$ for some neighbor $u$ of $v$, we utilize the mnemonics ``$\cn$,'' ``$\ncnI$,'' and ``$\ncnII$''
as references, respectively, to the cases of $v \in C_u$, $v \not \in C_u$ with $c_v=1$, and $v \not \in C_u$ with $c_v=2$.
These mnemonics are intended to facilitate the use of the probabilities calculated above (and also the ones we are about to
calculate) later in this section.

Let $u$ be a randomly chosen node and $v$ one of its neighbors in $G$ such that $v \in C_u$. Nodes $u$ and $v$ are then neighbors in $D$.
Let $b$ be the degree of $v$ in $G$. The probability that $\vert C_v \setminus \{u\} \vert = 0$ (i.e., $v$ chose $u$ and no other neighbor 
besides $u$), which we denote by $\pi^{(b)}_\cn(0)$, is the probability that all of $v$'s choices resulted in $u$, that is,
\begin{equation}
   \pi^{(b)}_\cn(0) = (1-\alpha) \frac{1}{b} + \alpha \frac{1}{b^2} = \frac{1 - \alpha + \alpha/b}{b}.
\end{equation}
Similarly, the probability that $\vert C_v \setminus \{u\} \vert = 1$, which we denote by $\pi^{(b)}_\cn(1)$, is
\begin{equation}
   \pi^{(b)}_\cn(1) = (1-\alpha)\left(\frac{b-1}{b}\right) + \alpha \left[3\left(\frac{b-1}{b^2}\right)\right] =
                  \left(\frac{b-1}{b}\right)(1-\alpha+3\alpha/b),
\end{equation}
which means that either $c_v=1$ and $u \not \in C_v$, or $c_v=2$ and either $v$ chose $u$ exactly once or
it did not but its choices were identical.
Finally, the probability that $\vert C_v \setminus \{u\} \vert = 2$, which we denote by $\pi^{(b)}_\cn(2)$,
is the probability that $c_v=2$ while $v$'s choices were distinct both from each other and from $u$, that is,
\begin{equation}
   \pi^{(b)}_\cn(2) = \alpha \left(\frac{b-1}{b}\right)\left(\frac{b-2}{b}\right).
\end{equation}
(It is worth noting that $\pi^{(b)}_\cn(0)$, $\pi^{(b)}_\cn(1)$, and $\pi^{(b)}_\cn(2)$ remain as calculated even without the condition that
$v \in C_u$, and in this case the use of ``$\cn$'' is pointless. We do insist on $v \in C_u$, however, because this is the context
in which the three probabilities are used in the sequel.)

In an analogous way, let us consider a randomly chosen node $u$ and a degree-$b$ neighbor $v$ of $u$ in $G$ such that $v \not \in C_u$. 
If this is the case, then $u$ and $v$ are neighbors in $D$ if and only if $u \in C_v$.
If $c_v=1$, then $u \in C_v$ if and only if $\vert C_v \setminus \{u\} \vert = 0$, and this happens with a probability that we
denote by $\pi^{(b)}_\ncnI(0)$ and is such that
\begin{equation}
   \pi^{(b)}_\ncnI(0) = 1/b.
\end{equation}
So $1 - \pi^{(b)}_\ncnI(0)$ is the probability that $u$ and $v$ are not neighbors in $D$, given that $v \not \in C_u$ and $c_v=1$.
If $c_v=2$, then the probability that $u \in C_v$ is $(2b-1)/b^2$, as we see from the fact that either $\vert C_v \setminus \{u\} \vert = 0$
or $\vert C_v \setminus \{u\} \vert = 1$ may happen. The probability of the former, denoted by $\pi^{(b)}_\ncnII(0)$, is given by
\begin{equation}
   \pi^{(b)}_\ncnII(0) = \frac{1}{b^2},
\end{equation}
while the probability of the latter, denoted by $\pi^{(b)}_\ncnII(1)$, refers to exactly one of $v$'s choices resulting in $u$. Thence
\begin{equation}
   \pi^{(b)}_\ncnII(1) = \frac{2b-2}{b^2}.
\end{equation}
So, given that $v \not \in C_u$ and $c_v=2$, $1 - \pi^{(b)}_\ncnII(0) - \pi^{(b)}_\ncnII(1)$ is the probability that $u$ and $v$ are not neighbors 
in $D$.

In the remainder of this section we analyze the efficacy of the uniform approach, concerning the performance requirements mentioned in 
Section~\ref{sec:introduction} when $D$ is used to disseminate $I$ from the originator.
Recall that we assume the limiting case of $n \to \infty$, 
so the probability that a finite-length cycle exists is negligible.
We also assume that both $G$ and $D$ are above their phase transitions (that is, both $\GCC{G}$ and $\GCC{D}$ almost surely exist)
and predicate our analysis upon the originator being a member of $\GCC{G}$.

\subsection{Number of nodes reached}

Let $\Pn$ be the ratio of the expected number of nodes reached when $I$ is disseminated on
$D$'s edges to the expected number of nodes in $\GCC{G}$.
Let also $\theta_G$ and $\theta_D$ be the fractions of $n$ corresponding to nodes inside $\GCC{G}$ and $\GCC{D}$, respectively. 
There are two cases to be considered. The first case is the one in which the originator  is a member of $\GCC{D}$, 
which occurs with probability $\theta_D/\theta_G$, this being also the ratio in this case. The second case corresponds to the
originator being outside $\GCC{D}$ and then the ratio is negligible. Therefore, $\Pn$ is
given by
\begin{equation}
   \Pn = \frac{\theta_D^2}{\theta_G^2}.
\end{equation}

Given a node $u$ and one of its neighbors, say $v$, we define the reach of $u$ through $v$ in $G$ as the set of nodes reachable
by a path in $G$ starting at $u$ whose first edge is $(u,v)$. We call $v$ a dead end with respect to $u$ in $G$ if the reach of $u$ 
through $v$ in $G$ is $o(n)$. Clearly, $v$ is a dead end with respect to $u$ in $G$, and this happens with probability denoted by $q$,
if and only if each of the other neighbors of $v$, in turn, is itself a dead end with respect to $v$ in $G$, and this happens
with probability $q$ for each of those neighbors as well.
If the degree of $v$ is $b$, then it is a dead end with respect to $u$ in $G$ with probability $q^{b-1}$.
And since the probability that $v$ has degree $b$ is given as in (\ref{eq:degneigh}), this leads to
\begin{equation}
   q=\sum_{b=1}^{n-1} \frac{b P_G(b)}{Z_G}q^{b-1}.
   \label{eq:q}
\end{equation}
Similarly, a randomly chosen node is not in $\GCC{G}$ if and only if each of its neighbors is a dead end with respect to it in $G$.
The expected fraction of nodes inside $\GCC{G}$ is then
\begin{equation}
   \theta_G = 1 - \sum_{a=0}^{n-1} P_G(a)q^a.
\end{equation}

The value of $\theta_D$ can be obtained in a similar, albeit more complex, way; it relies on definitions of reach and of dead-end nodes
in $D$ that are completely analogous to the ones in $G$.
Let $u$ be a randomly chosen node and $v$ one of its neighbors in $G$. Let also $q_\cn$, $q_\ncnI$, and $q_\ncnII$ be 
the conditional probabilities that $v$ is a dead end with respect to $u$ in $D$ given, respectively, that $v \in C_u$, that 
$v \not \in C_u$ with $c_v=1$, and that $v \not \in C_u$ with $c_v=2$. Regardless of which of these three conditions the case is, 
$v$ is a dead end with respect to $u$ in $D$ if and only if either $v$ is not a neighbor of $u$ in $D$ (which cannot happen under the first
condition), or it is but the reach of $u$ through $v$ in $D$ is $o(n)$, which means that all the other neighbors of $v$ in $G$ are themselves 
dead ends with respect to it in $D$. We may then write
\begin{eqnarray}
   \lefteqn{q_\cn = \sum_{b=1}^{n-1} \frac{bP_G(b)}{Z_G}\left[\pi^{(b)}_\cn(0) \left( \left(1-\alpha\right)q_\ncnI + \alpha q_\ncnII \right)^{b-1} \right.} \hspace{\eqIndent} \nonumber \\
                                        & &   \mbox{} + \pi^{(b)}_\cn(1) q_\cn \left( \left(1-\alpha\right)q_\ncnI + \alpha q_\ncnII \right)^{b-2} \nonumber \\
                                        & &   \mbox{} + \left.\pi^{(b)}_\cn(2) q_\cn^2 \left( \left(1-\alpha\right)q_\ncnI + \alpha q_\ncnII \right)^{b-3}\right],
   \label{eq:qcn}
\end{eqnarray}
\begin{equation}
   q_\ncnI = \sum_{b=1}^{n-1} \frac{bP_G(b)}{Z_G}\left[1 - \pi^{(b)}_\ncnI(0) + \pi^{(b)}_\ncnI(0) \left( \left(1-\alpha\right)q_\ncnI + \alpha q_\ncnII \right)^{b-1} \right],
   \label{eq:qncnI}
\end{equation}
and
\begin{eqnarray}
   \lefteqn{q_\ncnII = \sum_{b=1}^{n-1} \frac{bP_G(b)}{Z_G}\left[1 - \pi^{(b)}_\ncnII(0) - \pi^{(b)}_\ncnII(1) \right.} \hspace{\eqIndent} \nonumber \\
                                        & &   \mbox{} + \pi^{(b)}_\ncnII(0) \left( \left(1-\alpha\right)q_\ncnI + \alpha q_\ncnII \right)^{b-1} \nonumber \\
                                        & &   \mbox{} + \left.\pi^{(b)}_\ncnII(1) q_\cn \left( \left(1-\alpha\right)q_\ncnI + \alpha q_\ncnII \right)^{b-2}\right],
      \label{eq:qncnII}
\end{eqnarray}
referring back to the probabilities calculated in the introduction to Section~\ref{sec:analysis}.

Each of the expressions in (\ref{eq:qcn})--(\ref{eq:qncnII}) illustrates our use of those probabilities. We comment on (\ref{eq:qcn})
in detail and urge the reader to consider each of the others, as well as other expressions yet to come, in a similar light. In
(\ref{eq:qcn}), and for $k \in \{0,1,2\}$, $q_\cn^k\left((1-\alpha)q_\ncnI + \alpha q_\ncnII\right)^{b-1-k}$ gives the conditional
probability that all the $b-1$ neighbors of a degree-$b$ neighbor $v$ of $u$ that are not $u$ are dead ends with respect to $v$ in $D$;
the condition is that $\vert C_v \setminus \{u\}\vert = k$, which happens with probability $\pi^{(b)}_\cn(k)$. Thus $q_\cn^k$ is the
dead-end probability for the $k$ neighbors in $C_v \setminus \{u\}$, and $\left((1-\alpha)q_\ncnI + \alpha q_\ncnII\right)$ is the
dead-end probability for each of the
$b-1-k$ neighbors that are not in $C_v \cup \{u\}$.

Now, since a randomly chosen node is not in $\GCC{D}$ if and only if each of its neighbors in $G$ is a dead end with respect to it in $D$,
we have
\begin{eqnarray}
   \lefteqn{\theta_D = 1 - P_G(0) - \sum_{a=1}^{n-1} P_G(a) \left[ \pi^{(a)}_\rn(1) q_\cn \left( \left(1-\alpha\right)q_\ncnI + \alpha q_\ncnII \right)^{a-1} \right.} \hspace{\eqIndent} \nonumber \\
                                     & & \mbox{} + \left. \pi^{(a)}_\rn(2) q_\cn^2 \left( \left(1-\alpha\right)q_\ncnI + \alpha q_\ncnII \right)^{a-2} \right].
   \label{eq:thetaD}
\end{eqnarray}
Notice that, in (\ref{eq:thetaD}), $P_G(0)$ must be singled out of the sum in order to be taken into account, since $\pi^{(0)}_\rn(1)=\pi^{(0)}_\rn(2)=0$.

\subsection{Number of messages sent} \label{sec:pm}

We denote by $\Pm$ the ratio of the expected number of messages sent when disseminating $I$ through $D$'s edges to
the expected number of messages sent when disseminating $I$ through the edges of a spanning tree of $\GCC{G}$.
Let $Z_\GCC{D}$ be the average degree of nodes in $D$ conditioned upon membership in $\GCC{D}$.
Restricting our analysis to the case in which the originator is in $\GCC{D}$ (the other case leads to a negligible ratio),
which occurs with probability $\theta_D/\theta_G$,
the expected number of messages sent on the edges of $D$ is $n\theta_DZ_\GCC{D}$, while the expected number of messages sent on
the edges of the spanning tree of $\GCC{G}$ is $2\left(n\theta_G-1\right)$, thus leading to
\begin{equation}
   \Pm = \frac{\theta_D^2 n Z_\GCC{D}}{2\theta_G\left(n \theta_G -1\right)}.
\end{equation}

Now let $P_D(i \mid \GCC{D})$ be the probability that a randomly chosen node of $\GCC{D}$ has degree $i$ in $D$. We have
\begin{equation}
   Z_\GCC{D} = \sum_{i=0}^{n-1} i P_D(i \mid \GCC{D}).
\end{equation}
If also we let $P_D(i\mid a, \GCC{D})$ be the conditional probability that a randomly chosen node of $\GCC{D}$ has degree $i$ in $D$ given
that it has degree $a$ in $G$, then $P_D(i \mid \GCC{D})$ can be written as
\begin{equation}
   P_D(i \mid \GCC{D}) = \sum_{a=i}^{n-1} P_D(i \mid a, \GCC{D}) P_G(a\mid \GCC{D}),
\end{equation}
where $P_G(a\mid \GCC{D})$ is the probability that the node has degree $a$ in $G$.
We henceforth approximate $P_D(i \mid a, \GCC{D})$ by $P_D(i \mid a)$, which is the probability that a randomly
chosen, degree-$a$ node of $G$ has degree $i$ in $D$ (more on this approximation in Section~\ref{sec:node_degree}).
Also, if $u$ is this node and $v$ one of its neighbors in $G$ such that $v \not \in C_u$, 
then the probability that $u \in C_v$, denoted by $r$, is given by
\begin{equation}
   r = \sum_{b=1}^{n-1} \frac{b P_G(b)}{Z_G} \left[\left(1-\alpha\right)\pi^{(b)}_\ncnI(0) + \alpha\left(\pi^{(b)}_\ncnII(0) + \pi^{(b)}_\ncnII(1)\right)\right]
   \label{eq:r}
\end{equation}
and yields
\begin{equation}
   P_D(i \mid a) = \pi^{(a)}_\rn(1) \binom{a-1}{i-1} r^{i-1} \left(1-r\right)^{a-i} + \pi^{(a)}_\rn(2) \binom{a-2}{i-2} r^{i-2} \left(1-r\right)^{a-i}.
   \label{eq:pia}
\end{equation}

Using Bayes' rule to rewrite $P_G(a \mid \GCC{D})$ as
\begin{equation}
   P_G(a\mid \GCC{D}) = \frac{P_D(\GCC{D} \mid a) P_G(a)}{P_D(\GCC{D})},
   \label{pagccd}
\end{equation}
where $P_D(\GCC{D})=\theta_D$ and
\begin{eqnarray}
   \lefteqn{P_D(\GCC{D} \mid a) = 1 - \pi^{(a)}_\rn(1) q_\cn \left( \left(1-\alpha\right)q_\ncnI + \alpha q_\ncnII \right)^{a-1}} \hspace{\eqIndent} \nonumber \\
                                & & \mbox{} - \pi^{(a)}_\rn(2) q_\cn^2 \left( \left(1-\alpha\right)q_\ncnI + \alpha q_\ncnII \right)^{a-2},
                                \label{eq:pgccda}
\end{eqnarray}
leads, finally, to
\begin{eqnarray}
   Z_\GCC{D} & = & \sum_{i=0}^{n-1} i \sum_{a=i}^{n-1}  P_D(i\mid a) \frac{P_D(\GCC{D} \mid a) P_G(a)}{\theta_D} \nonumber \\
             & = & \sum_{a=0}^{n-1} \frac{P_D(\GCC{D} \mid a) P_G(a)}{\theta_D} \sum_{i=0}^{a} i P_D(i\mid a),
   \label{eq:zgccd}
\end{eqnarray}
where
\begin{equation}
   \sum_{i=0}^{a} i P_D(i\mid a) = \pi^{(a)}_\rn(1) \left(1 + (a-1)r\right) + \pi^{(a)}_\rn(2) \left(2 + (a-2)r\right).
   \label{eq:epia}
\end{equation}

\subsection{Node degree} \label{sec:node_degree}

Let us now consider the expected degree of $D$ conditioned upon membership in $\GCC{G}$, which we denote by $Z_{D,\GCC{G}}$.
Let $P_D(i \mid \GCC{G})$ be the probability that a randomly chosen node of $\GCC{G}$ has degree $i$ in $D$.
Then $Z_{D,\GCC{G}}$ is clearly given by
\begin{equation}
   Z_{D,\GCC{G}} = \sum_{i=0}^{n-1} i P_D(i \mid \GCC{G}),
   \label{eq:zdgccg1}
\end{equation}
where
\begin{equation}
P_D(i \mid \GCC{G}) = \sum_{a=i}^{n-1} P_D(i \mid a, \GCC{G}) P_G(a \mid \GCC{G}),
\end{equation}
$P_D(i \mid a, \GCC{G})$ being the probability that a randomly chosen, degree-$a$ node of $\GCC{G}$ has degree
$i$ in $D$. This lets (\ref{eq:zdgccg1}) be rewritten as
\begin{eqnarray}
   Z_{D,\GCC{G}} & = & \sum_{i=0}^{n-1} i \sum_{a=i}^{n-1} P_D(i \mid a, \GCC{G}) P_G(a \mid \GCC{G}) \nonumber \\
                 & = & \sum_{a=0}^{n-1} P_G(a \mid \GCC{G}) \sum_{i=0}^{a} i P_D(i \mid a, \GCC{G}),
\end{eqnarray}
which, using Bayes' rule to write
\begin{equation}
   P_G(a \mid \GCC{G}) = \frac{P_G(\GCC{G} \mid a)P_G(a)}{P_G(\GCC{G})} = \frac{\left(1 - q^a\right)P_G(a)}{\theta_G}
   \label{eq:pagccg}
\end{equation}
(cf.\ (\ref{eq:q})), yields
\begin{equation}
  Z_{D,\GCC{G}} = \sum_{a=0}^{n-1} \frac{\left(1 - q^a\right)P_G(a)}{\theta_G} \sum_{i=0}^{a} i P_D(i \mid a, \GCC{G}).
  \label{eq:zdgccg2}
\end{equation}

One possibility now would be to proceed similarly to what we did in Section~\ref{sec:pm} and approximate $P_D(i \mid a, \GCC{G})$ by $P_D(i \mid a)$.
This would immediately let us use (\ref{eq:epia}) in (\ref{eq:zdgccg2}) and be done. However, we know from early experiments like the ones to be discussed in
Section~\ref{sec:simulation} that, unlike the case of Section~\ref{sec:pm}, this sometimes yields an agreement between analytical prediction and
simulation that is not satisfactory. In the present case, then, we look more closely at the nature of $P_D(i \mid a, \GCC{G})$ and first notice that
(\ref{eq:pia})---and consequently (\ref{eq:epia}) as well---can essentially be used to express $P_D(i \mid a, \GCC{G})$, provided the probability $r$
appearing in it is made to depend on $a$ and on the degree-$a$ node's membership in $\GCC{G}$. This is to be taken in opposition to the expression for
$r$ in (\ref{eq:r}), but clearly all that needs to be changed in (\ref{eq:r}) to make the dependencies manifest is to replace $bP_G(b)/Z_G$ as the
probability that a given neighbor of a degree-$a$ node has degree $b$. The reason why this is so is that, given a degree-$a$ node's membership in
$\GCC{G}$, that probability is no longer independent from $a$ (as we know from our comment following (\ref{eq:criterion2}), the existence of $\GCC{G}$ is
related to nodes' neighbors' degrees).

Let $p$ be the probability that we seek for use in place of $bP_G(b)/Z_G$. That is, $p$ is the probability that a given neighbor of a degree-$a$ node of
$\GCC{G}$ has degree $b$. The probability that a degree-$a$ node is in $\GCC{G}$ is $1-q^a$ (all its $a$ neighbors must otherwise be dead ends with respect
to it in $G$), so the probability that a degree-$a$ node is a member of $\GCC{G}$ and moreover a given neighbor of it has degree $b$ is $p(1-q^a)$.
But this latter probability can also be expressed as $\left(bP_G(b)/Z_G\right)(1-q^{a+b-2})$,
since the rightmost factor is the probability of the condition that a pair of neighbors (one of degree $a$, the other of
degree $b$) is in $\GCC{G}$, and furthermore $bP_G(b)/Z_G$, under that condition, continues to give the probability that a given neighbor of a degree-$a$
node has degree $b$. We may then write
\begin{equation}
p(1-q^a) = \frac{bP_G(b)}{Z_G}(1-q^{a+b-2}),
\end{equation}
from which it follows that
\begin{equation}
p = \frac{bP_G(b)}{Z_G}\left(\frac{1-q^{a+b-2}}{1-q^a}\right).
\end{equation}

We note, finally, that an analogous but more complicated development could also be used to avoid the approximation of $P_D(i \mid a, \GCC{D})$ by
$P_D(i \mid a)$ made in Section~\ref{sec:pm}. As explained, however, that would only add needless detail.

\subsection{Path length} \label{sec:pt}

Let $\Pt$ be the ratio of the expected path length from the originator in $D$ to the expected path length from the originator in $G$. We
may again restrict our analysis to the case in which the originator is a member of $\GCC{D}$, which occurs with probability $\theta_D/\theta_G$.
Denoting by $L_\GCC{G}$ and $L_\GCC{D}$ the average path lengths from the originator in $G$ and in $D$, respectively,
we have
\begin{equation}
   \Pt = \frac{\theta_D L_\GCC{D}}{\theta_G L_\GCC{G}}.
\end{equation}

By (\ref{eq:pagccg}), the average degree in $G$ of the nodes inside $\GCC{G}$, denoted by $Z_\GCC{G}$, is
\begin{equation}
   Z_\GCC{G} = \sum_{a=0}^{n-1} \frac{a \left( 1- q^a \right) P_G(a)}{\theta_G}.
\end{equation}
Now let two nodes be called $\ell$-neighbors in a graph, for $\ell \geq 1$, when the distance between them in the graph is $\ell$ (the case of
$\ell=1$ is simply the case of neighbors in the graph). Given a randomly chosen node $u$ in $\GCC{G}$ and one of its neighbors,
say $v$, the expected number of $v$'s other neighbors (i.e., excluding $u$), denoted by $\rho$, is\footnote{The use of $bP_G(b)/Z_G$ in
(\ref{eq:rho}) and later in (\ref{eq:rncnI})--(\ref{eq:tncnII}) is in principle subject to the same corrections explained at the end of
Section~\ref{sec:node_degree}. However, introducing those corrections in the present context not only seems unnecessary given the computational
results to be discussed in Section~\ref{sec:simulation}, but also would lead to much more complicated (and probably insoluble)
versions of (\ref{eq:sum1}) and (\ref{eq:sum2}).}
\begin{equation}
   \rho = \sum_{b=1}^{n-1} \frac{b P_G(b)}{Z_G} (b-1),
   \label{eq:rho}
\end{equation}
and then the expected number of $u$'s $2$-neighbors in $G$, which we denote by $\ZGCC{2}{G}$, is
\begin{equation}
   \ZGCC{2}{G} = Z_\GCC{G} \rho.
\end{equation}
In general, the expected number of $u$'s $\ell$-neighbors in $G$, denoted by $\ZGCC{\ell}{G}$, is
\begin{equation}
   \ZGCC{\ell}{G} = \ZGCC{\ell-1}{G} \rho = Z_\GCC{G} \rho^{\ell-1}.
   \label{eq:zrgccg}
\end{equation}

We can then obtain an approximation for $L_\GCC{G}$ by summing the values of $\ZGCC{\ell}{G}$ from $\ell=1$ up until the sum becomes equal to
the expected number of nodes inside $\GCC{G}$ minus one (to account for node $u$). Thus
\begin{equation}
   \sum_{\ell=1}^{L_\GCC{G}} \ZGCC{\ell}{G} = n \theta_G - 1,
   \label{eq:sum1}
\end{equation}
which yields
\begin{equation}
   L_\GCC{G} = \frac{\ln\left[ \left( \frac{n\theta_G-1}{Z_\GCC{G}} \right) \left( \rho -1 \right) + 1\right]}{\ln \rho}.
\end{equation}

Let us now turn to obtaining the value of $L_\GCC{D}$. Consider a randomly chosen node $u$ of $\GCC{D}$ and a neighbor $v$ of $u$ in $G$.
If $v \not \in C_u$, then let $r_\ncnI$ be the conditional probability that $u \in C_v$ given that $c_v=1$. We have
\begin{equation}
   r_\ncnI = \sum_{b=1}^{n-1} \frac{b P_G(b)}{Z_G} \pi^{(b)}_\ncnI(0).
   \label{eq:rncnI}
\end{equation}
Likewise, if $r_\ncnII$ is the conditional probability that $u \in C_v$ given that $c_v=2$, then
\begin{equation}
   r_\ncnII = \sum_{b=1}^{n-1} \frac{b P_G(b)}{Z_G} \left(\pi^{(b)}_\ncnII(0)+\pi^{(b)}_\ncnII(1) \right).
   \label{eq:rncnII}
\end{equation}

Calculating the expected number of other neighbors of a neighbor $v$ of $u$ requires three cases to be considered. The first
case is the case of $v \in C_u$, and then the expected number of $u$'s $2$-neighbors in $D$ that are reachable from $v$ is 
$t_\cn(1,r_\ncnI,r_\ncnII)$, where
\begin{eqnarray}
   \lefteqn{t_\cn(x,y,z) = \sum_{b=1}^{n-1} \frac{b P_G(b)}{Z_G} \left\{
      \pi_\cn^{(b)}(0) \left(b-1\right)\left(\left(1-\alpha\right)y + \alpha z\right) \right.} \hspace{\eqIndent} \nonumber \\
      & & \mbox{} + \pi_\cn^{(b)}(1) \left[x + \left(b-2\right)\left(\left(1-\alpha\right)y + \alpha z\right)\right] \nonumber \\
      & & \mbox{} + \left.\pi_\cn^{(b)}(2) \left[2x + \left(b-3\right)\left(\left(1-\alpha\right)y + \alpha z\right)\right]\right\}.
      \label{eq:tcn}
\end{eqnarray}

In the second case, $v \not \in C_u$ with $c_v=1$. We similarly let
\begin{equation}
   t_\ncnI(y,z) = \sum_{b=1}^{n-1} \frac{b P_G(b)}{Z_G} \left[\pi_\ncnI^{(b)}(0) \left(b-1\right)\left(\left(1-\alpha\right)y + \alpha z\right)\right],
   \label{eq:tncnI}
\end{equation}
and then the expected number of $u$'s $2$-neighbors in $D$ that are reachable from $v$ is $t_\ncnI(r_\ncnI,r_\ncnII)$.

The third and final case is that of $v \not \in C_u$ with $c_v=2$. As in the previous two cases, we let
\begin{eqnarray}
   \lefteqn{t_\ncnII(x,y,z) = \sum_{b=1}^{n-1} \frac{b P_G(b)}{Z_G} \left\{
      \pi_\ncnII^{(b)}(0) \left(b-1\right)\left(\left(1-\alpha\right)y + \alpha z\right) \right.} \hspace{\eqIndent} \nonumber \\
      & & \mbox{} + \left. \pi_\ncnII^{(b)}(1) \left[x + \left(b-2\right)\left(\left(1-\alpha\right)y + \alpha z\right)\right] \right\},
   \label{eq:tncnII}
\end{eqnarray}
which yields the expected number of $u$'s $2$-neighbors in $D$ that are reachable from $v$ as $t_\ncnII(1,r_\ncnI, r_\ncnII)$.

Now, for $u$ a randomly chosen node in $\GCC{D}$, and recalling (\ref{pagccd}), letting
\begin{eqnarray}
   \lefteqn{t_\rn(x,y,z) = \sum_{a=0}^{n-1} \frac{P_D(\GCC{D} \mid a) P_G(a) }{\theta_D}} \hspace{\eqIndent}\nonumber \\
      & & \left\{\pi_\rn^{(a)}(1) \left[x + \left(a-1\right)\left(\left(1-\alpha\right)y + \alpha z\right)\right]\right. \nonumber \\
      & & \mbox{}+\left.\pi_\rn^{(a)}(2) \left[2x + \left(a-2\right)\left(\left(1-\alpha\right)y + \alpha z\right)\right]\right\}
   \label{eq:trn}
\end{eqnarray}
allows the expected degree of $u$ to be expressed as $t_\rn(1,r_\ncnI,r_\ncnII)$, and similarly the expected number of $2$-neighbors of $u$
as
\begin{equation}
   t_\rn(t_\cn(1,r_\ncnI,r_\ncnII),t_\ncnI(r_\ncnI,r_\ncnII),t_\ncnII(1,r_\ncnI,r_\ncnII)).
   \label{eq:z2d}
\end{equation}

For simplicity's sake, let
$\beta_\cn^\cn$, $\beta_\cn^\ncnI$, $\beta_\cn^\ncnII$,
$\beta_\ncnI^\ncnI$, $\beta_\ncnI^\ncnII$,
$\beta_\ncnII^\cn$, $\beta_\ncnII^\ncnI$, $\beta_\ncnII^\ncnII$,
$\beta_\rn^\cn$, $\beta_\rn^\ncnI$, and $\beta_\rn^\ncnII$
be such that
(\ref{eq:tcn}), (\ref{eq:tncnI}), (\ref{eq:tncnII}), and (\ref{eq:trn}) can, respectively, be rewritten as
\begin{equation}
   t_\cn(x,y,z)    = \beta_\cn^\cn x    + \beta_\cn^\ncnI y    + \beta_\cn^\ncnII z,
\end{equation}
\begin{equation}
   t_\ncnI(y,z)    = \beta_\ncnI^\ncnI y  + \beta_\ncnI^\ncnII z,
\end{equation}
\begin{equation}
   t_\ncnII(x,y,z) = \beta_\ncnII^\cn x + \beta_\ncnII^\ncnI y + \beta_\ncnII^\ncnII z,
\end{equation}
and
\begin{equation}
   t_\rn(x,y,z)    = \beta_\rn^\cn x    + \beta_\rn^\ncnI y    + \beta_\rn^\ncnII z.
\end{equation}
Then, introducing the row vector
$A=\left[\begin{array}{ccc}
\beta_\rn^\cn & \beta_\rn^\ncnI & \beta_\rn^\ncnII
\end{array}\right]$ and the matrix
$$
   \Beta=\left[\begin{array}{ccc}
      \beta_\cn^\cn    & \beta_\cn^\ncnI    & \beta_\cn^\ncnII \\
      0                & \beta_\ncnI^\ncnI  & \beta_\ncnI^\ncnII \\
      \beta_\ncnII^\cn & \beta_\ncnII^\ncnI & \beta_\ncnII^\ncnII
   \end{array}\right],
$$
we have that the expected number of $\ell$-neighbors of a randomly chosen node in $\GCC{D}$, which for $\ell \geq 1$ we denote by 
$\ZGCC{\ell}{D}$, is
\begin{equation}
   \ZGCC{\ell}{D}= A\Beta^{\ell-1}\left[ \begin{array}{c}
      1 \\ r_\ncnI \\ r_\ncnII
   \end{array}\right].
   \label{eq:zrgccd}
\end{equation}
(The reader should check that (\ref{eq:zrgccd}) yields the $Z_\GCC{D}$ of (\ref{eq:zgccd}) for $\ell=1$, and also that it becomes 
(\ref{eq:z2d}) for $\ell=2$.)

We are now left with the task of finally obtaining $L_\GCC{D}$. This can only be achieved numerically, and to this end we resort to
the eigenvalues (say $\lambda_1$, $\lambda_2$, and $\lambda_3$, which we assume are all distinct\footnote{This has proven true in all the experimental
scenarios of Section~\ref{sec:simulation}, so we dwell on the matter no further.}) and corresponding eigenvectors
($v_1$, $v_2$, and $v_3$) of $\Beta$.
If $V$ is the matrix whose columns are $v_1$, $v_2$, and $v_3$, and $\Lambda$ the matrix having diagonal elements $\lambda_1$, 
$\lambda_2$, and $\lambda_3$ with $0$'s everywhere else, then $\Beta$ can be diagonalized into $\Lambda$ via
\begin{equation}
   \Lambda = V^{-1} \Beta V,
\end{equation}
which can be equivalently expressed as
\begin{equation}
   \Beta = V \Lambda V^{-1}
\end{equation}
and used to obtain the $\Beta^{\ell-1}$ of (\ref{eq:zrgccd}) as $\Beta^{\ell-1} = V \Lambda^{\ell-1} V^{-1}$ \cite{friedberg2003}.

Proceeding in a manner analogous to the one that led to (\ref{eq:sum1}), we can then obtain $L_\GCC{D}$ by numerically solving the
equation
\begin{equation}
   \sum_{\ell=1}^{L_\GCC{D}} \ZGCC{\ell}{D} = n \theta_D - 1,
   \label{eq:sum2}
\end{equation}
which, by (\ref{eq:zrgccd}), is equivalent to
\begin{equation}
   A V \left[\begin{array}{ccc}
      \frac{\lambda_1^{L_\GCC{D}}-1}{\lambda_1-1} &
      0                                           &
      0                                           \\
      0                                           &
      \frac{\lambda_2^{L_\GCC{D}}-1}{\lambda_2-1} &
      0                                           \\
      0                                           &
      0                                           &
      \frac{\lambda_3^{L_\GCC{D}}-1}{\lambda_3-1}
   \end{array}\right] V^{-1} \left[ \begin{array}{c}
      1 \\ r_\ncnI \\ r_\ncnII
   \end{array}\right] = n \theta_D - 1.
\end{equation}

\section{Simulation results} \label{sec:simulation}

In addition to our mathematical analysis of Section~\ref{sec:analysis}, and seeking to validate it experimentally, we have carried out 
simulations of the uniform approach to the construction of $D$. Also, and notwithstanding the fact that our analysis has not included
the degree-based approach, we have extended our simulations to cover it as well. Each of our simulations is based on disseminating
$I$ on random graphs, which are always generated to be above the corresponding phase transition. This allows the simulation to be 
constrained to operate within the graph's largest connected component, which almost surely is a giant connected component.

We have considered two random-graph models.
The first model corresponds to the classical model of Erd\H{o}s and
Rényi \cite{erdos1959}, in which $G$ is constructed on $n$ nodes by letting each of the possible $n(n-1)/2$ edges exist with
constant probability $z/(n-1)$ for $0 < z \leq n-1$. As a result, $G$ has node degrees distributed according to
a Poisson distribution, that is, the probability that a node has degree $a$ is $P_G(a)=e^{-z}z^a/a!$ \cite{bollobas2002}.
For Poisson-distributed node degrees, it follows from (\ref{eq:criterion}) that the graph is above the phase
transition if and only if $z > 1$, since $Z_G=z$ and $\mean{K_G^2}=z^2+z$.

We have concentrated on analyzing the behavior of $\Pn$, $\Pm$, $Z_{D,\GCC{G}}$, and $\Pt$ for $1 \leq z \leq 10$ and
$\alpha=0.10,0.25,0.50,0.75,1.00$. To this end, and for each value of $z$, we generated $300$ random graphs with $n=10000$ nodes,
and then constructed two instances of $D$ for each value of $\alpha$, one following the uniform approach and the other the degree-based
approach. On each of the instances, we then conducted $1000$ disseminations by randomly choosing an originator from among
the nodes in the largest connected component of $G$. At the end, we averaged the quantities of interest overall to obtain
$\Pn$, $\Pm$, $Z_{D,\GCC{G}}$, and $\Pt$. (Note that obtaining $Z_{D,\GCC{G}}$ does not depend on any dissemination, but rather only
on the available $G$ and $D$ instances. The same is in principle also true of $\Pt$, but we simulate disseminations by
breadth-first search from the originator and $\Pt$ can then be obtained along the way.)

\begin{figure*}[p]
   \centering

   \begin{tabular}{cc}
   \includegraphics[width=\graphWidth]{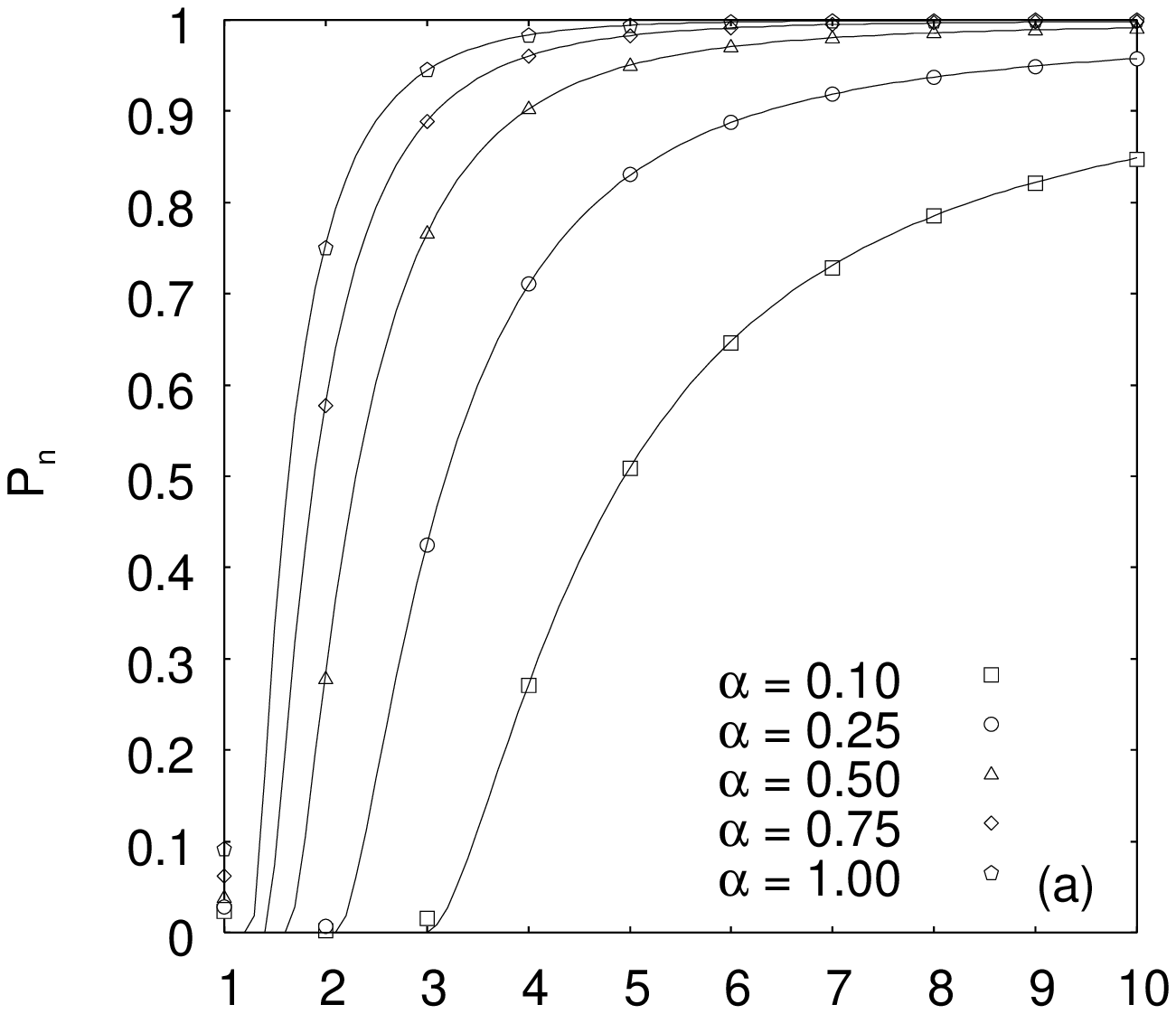}  &
   \includegraphics[width=\graphWidth]{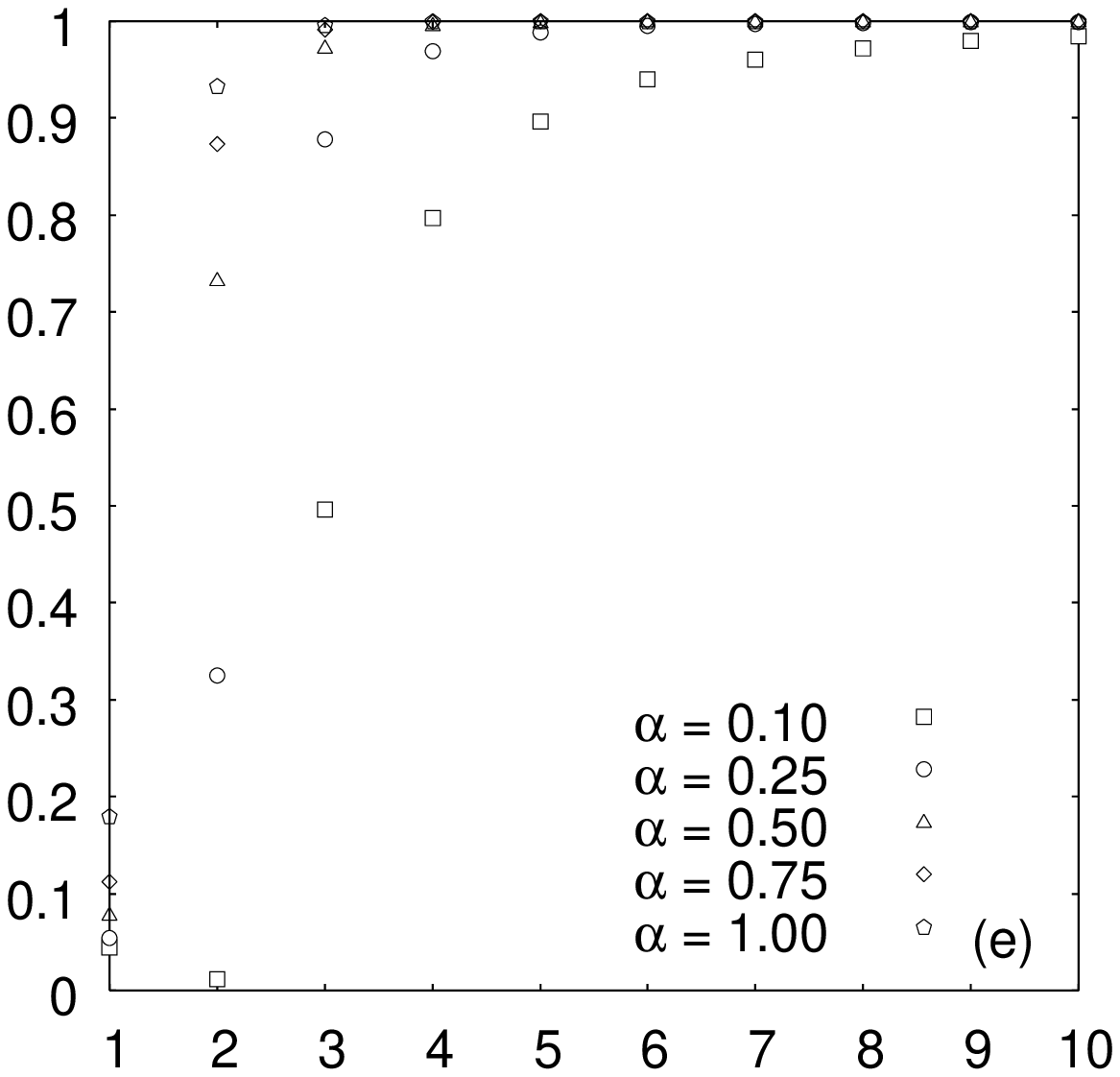}  \\
   \includegraphics[width=\graphWidth]{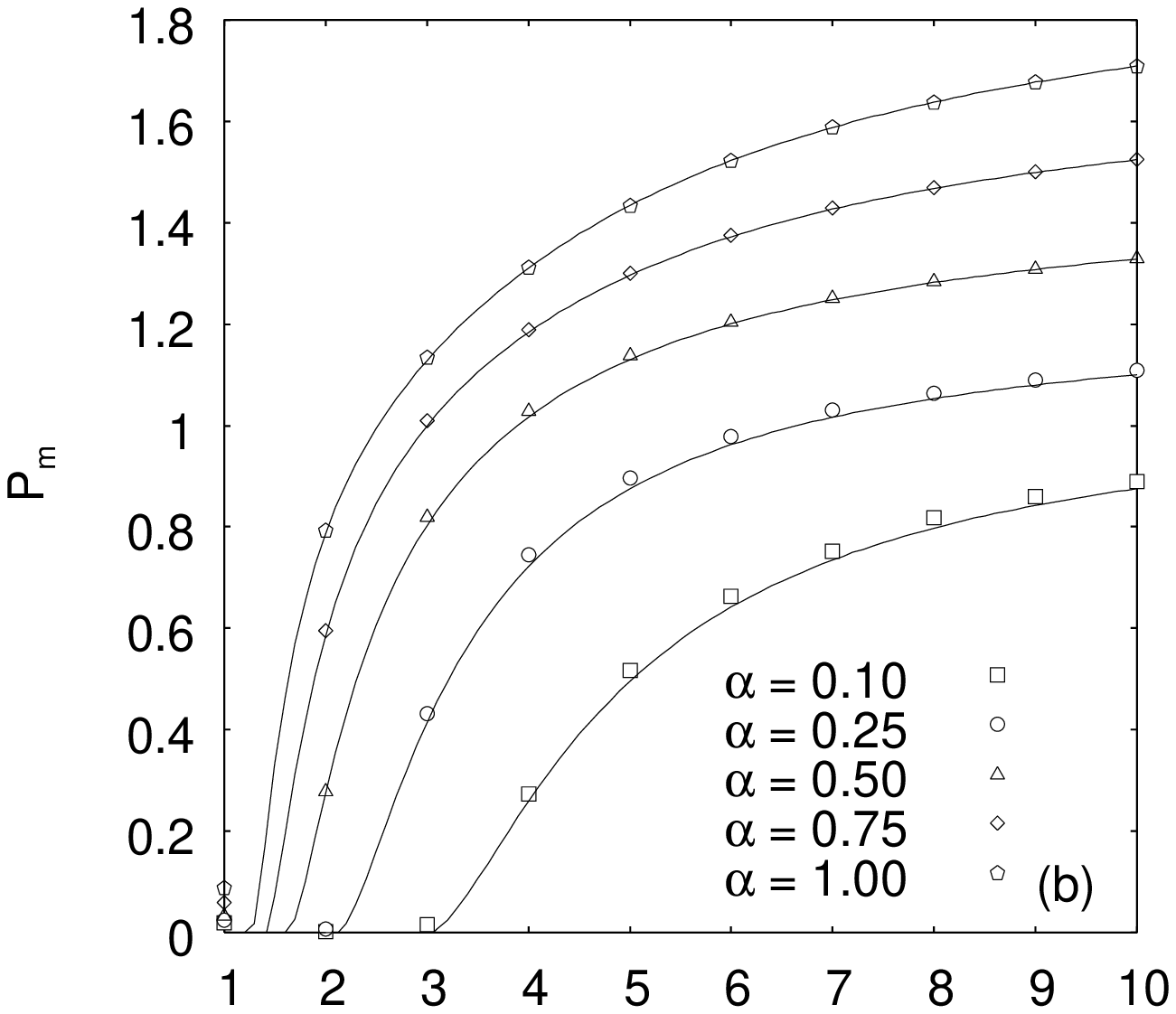}  &
   \includegraphics[width=\graphWidth]{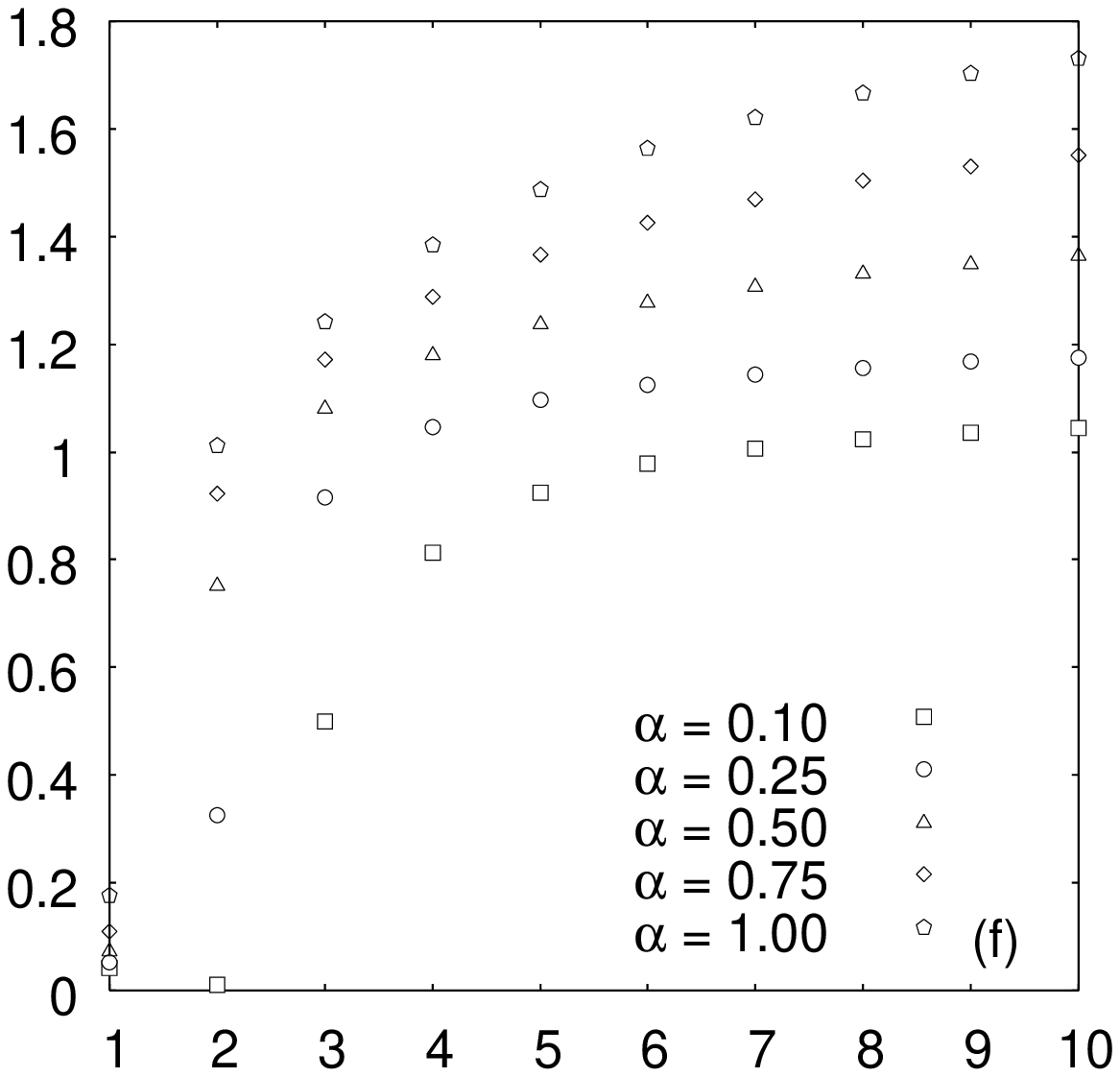}  \\
   \includegraphics[width=\graphWidth]{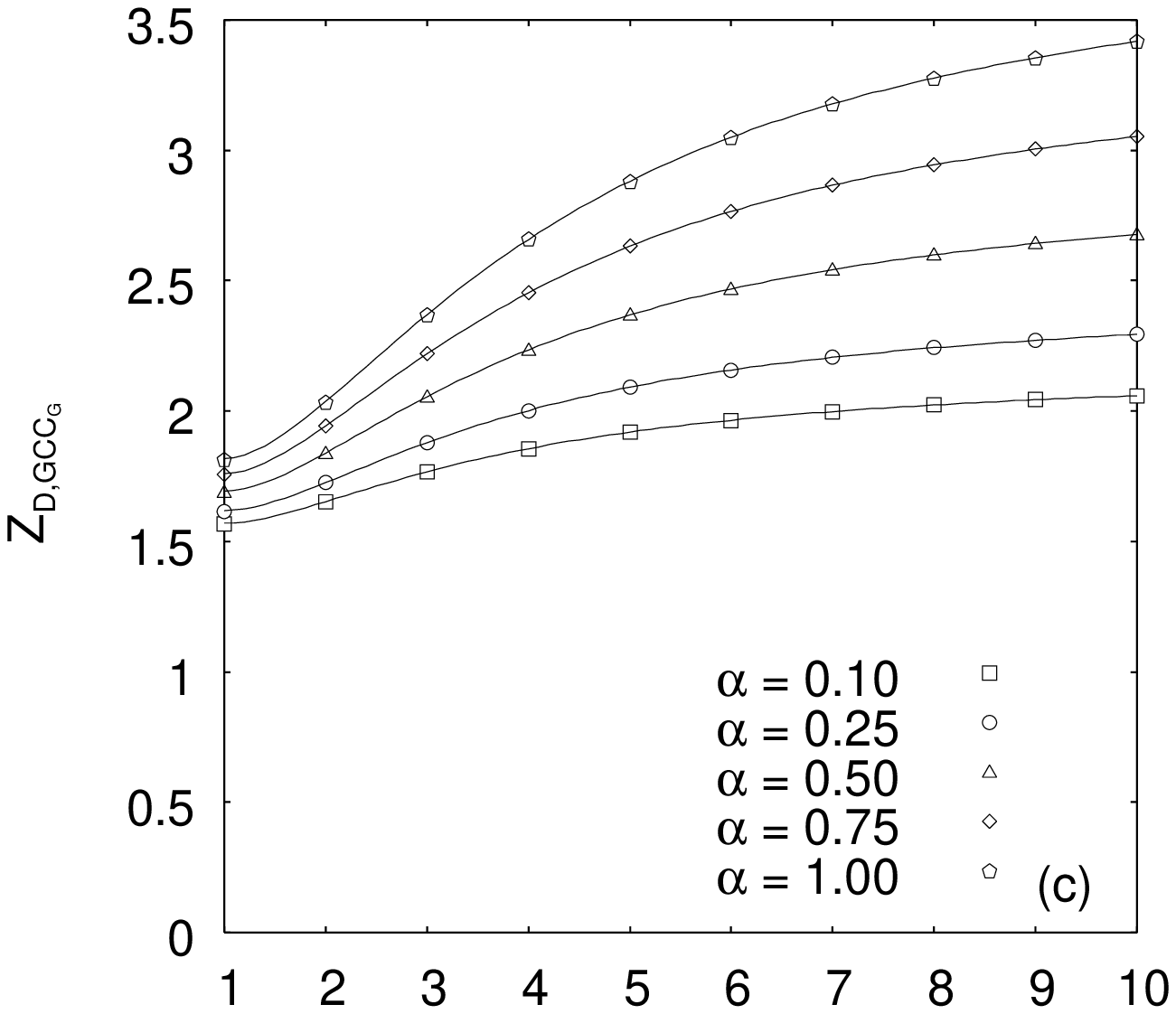}  &
   \includegraphics[width=\graphWidth]{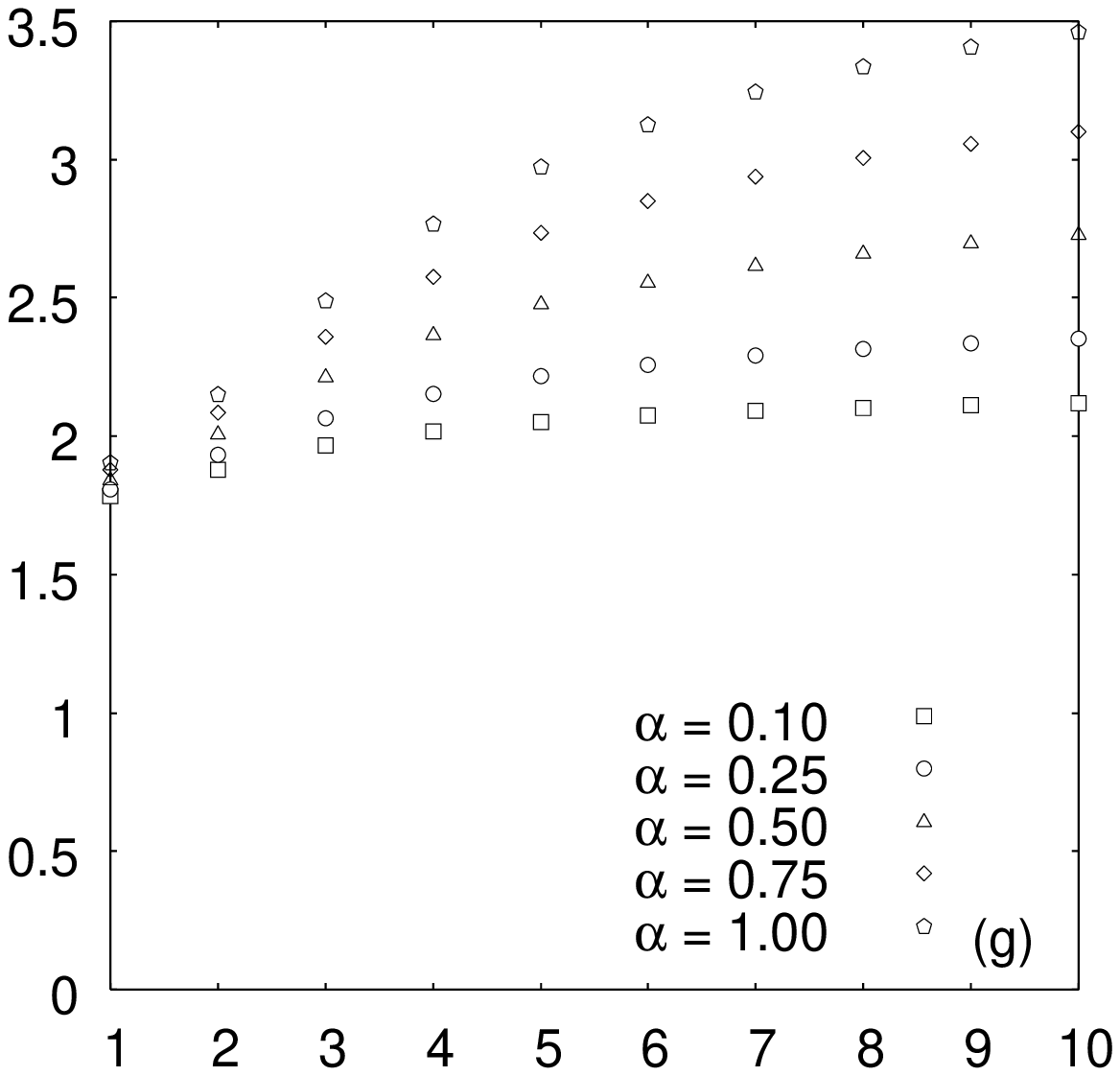}  \\
   \includegraphics[width=\graphWidth]{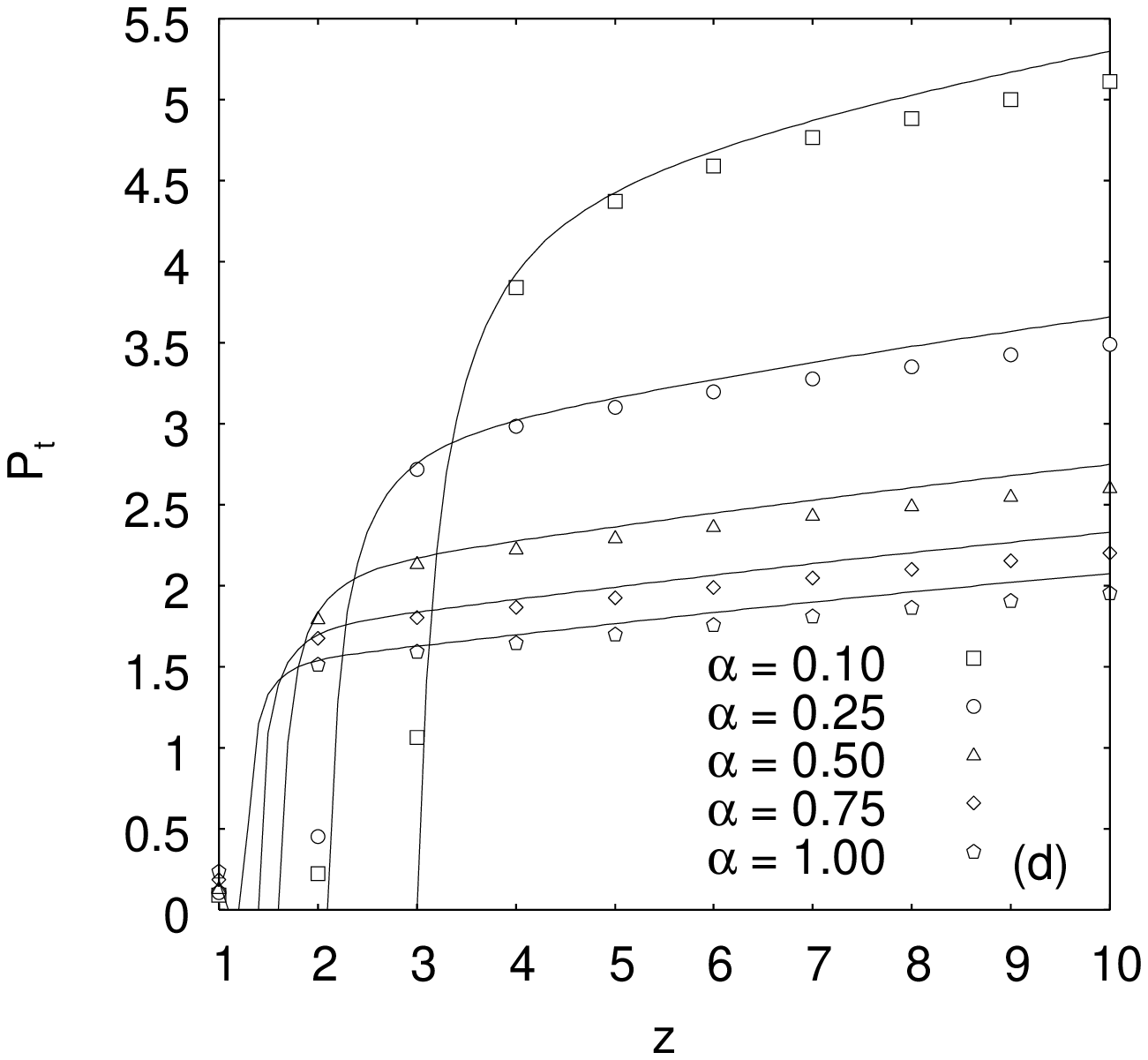} &
   \includegraphics[width=\graphWidth]{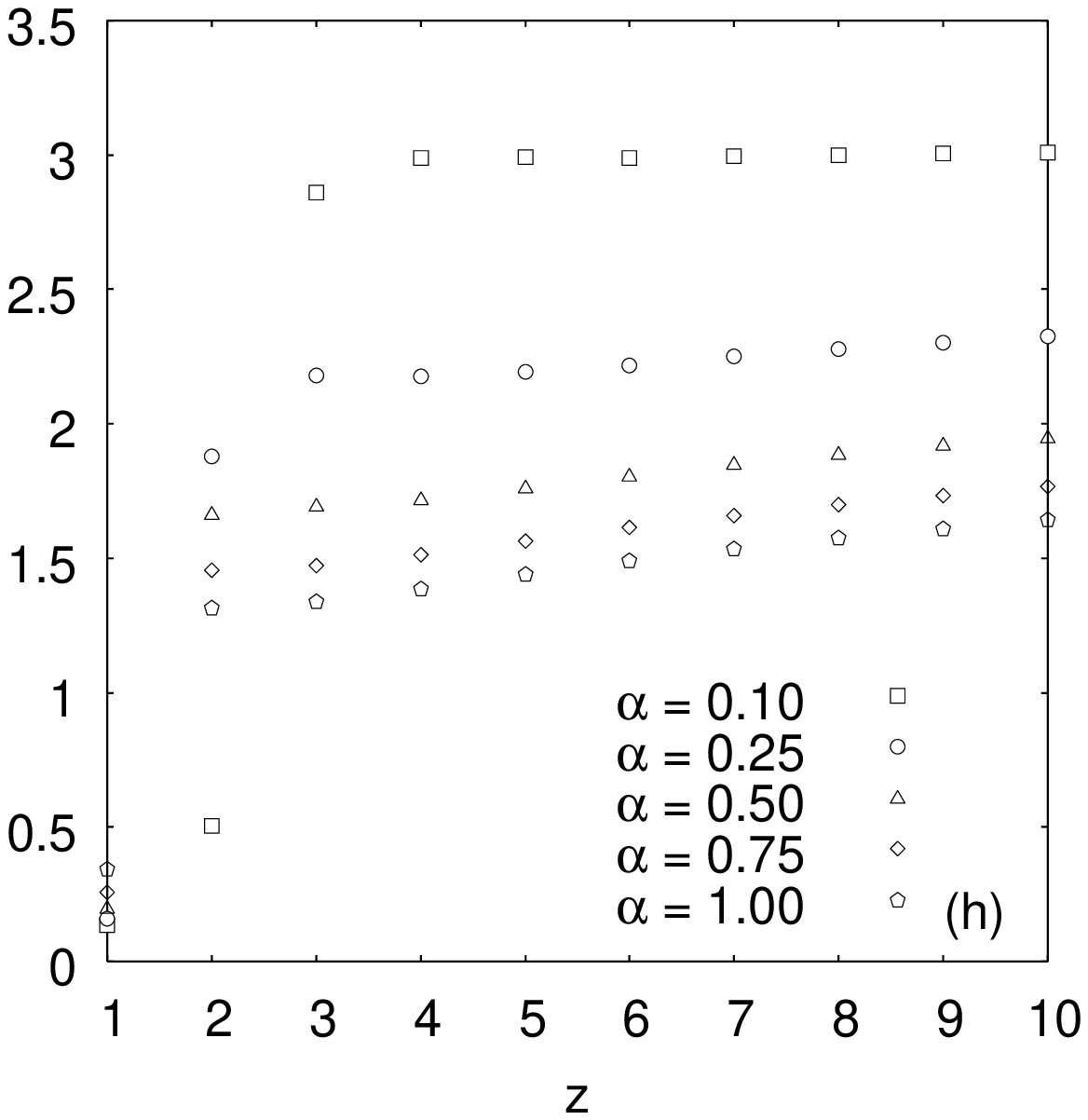}
   \end{tabular}

   \caption{Simulation results for the uniform (a--d) and the degree-based (e--h) approach on random graphs having
   Poisson-distributed node degrees. The plots show $\Pn$ (a, e), $\Pm$ (b, f), $Z_{D,\GCC{G}}$ (c, g), and $\Pt$ (d, h) for
   $\alpha=0.10,0.25,0.50,0.75,1.00$. Solid lines give the analytical predictions of Section~\ref{sec:analysis}.}
   \label{fig:sim_poisson}
\end{figure*}

Figure~\ref{fig:sim_poisson} shows simulation results for random graphs having Poisson-distributed node degrees.
Parts (a--d), concerning
the uniform approach, show an excellent agreement between analytical and simulation results, with only a
slight deviation in part (d), which is in all likelihood to be attributed to the approximations made in Section~\ref{sec:pt}. 
The plots for $\Pn$ (Figure~\ref{fig:sim_poisson}(a, e)) evidence the expected superiority of the degree-based approach over 
the uniform approach, since in the former case $\Pn$ approaches $1$ rapidly as $z$ is increased, more or less regardless of $\alpha$ 
(in the uniform approach, this only seems to happen for $z < 10$ when $\alpha \geq 0.50$). A closer examination of the data for 
the degree-based approach, say for $z=5$ and $\alpha=0.50$, reveals $\Pn \approx 0.998$, $\Pm \approx 1.24$, $Z_{D,\GCC{G}} \approx 2.48$, 
and $\Pt \approx 1.76$. What this means is that, using roughly $1.24$ times as many edges as a spanning tree and paths that, on average, are
greater than those of $G$ by a factor of only $1.76$, the dissemination subgraph reaches almost all the nodes of the network
while having a relatively low average node degree. 
Comparing the two approaches, it is curious to note that the plots of $\Pm$ (Figure~\ref{fig:sim_poisson}(b, f)) are very similar to
each other, the same holding for those of $Z_{D,\GCC{G}}$ (Figure~\ref{fig:sim_poisson}(c, g)), which indicates  that the number of 
edges in the dissemination subgraph is quite independent of whether one approach is used or the other. However, the difference between
the $\Pn$ plots demonstrates that the choices made by the nodes in the degree-based approach somehow lead the edges to end up deployed in
such a manner as to favor the connectedness of the dissemination subgraph strongly.

The other random-graph model we have considered is the one in which node degrees are distributed according to a power law. The probability
that a node in $G$ has degree $a$ is in this case, and for $n \to \infty$, given by $P_G(a)=a^{-\tau}/\zeta(\tau)$, where $\tau > 1$ is a
parameter and $\zeta(x)$ is the Riemann zeta function \cite{yan2002}, that is, $\zeta(x)=\sum_{y=1}^\infty y^{-x}$.
Then we have $Z_G=\zeta(\tau-1)/\zeta(\tau)$ and $\mean{K_G^2}=\zeta(\tau-2)/\zeta(\tau)$, so solving (\ref{eq:criterion}) numerically
yields $\tau < 3.47$ as the condition for $G$ to be above the phase transition. We have performed simulations for $2 \leq \tau \leq 3$
in the same way as we did for the Poisson case.

Random graphs with degrees thus distributed
can be generated in two phases. First the degrees $a_1, a_2, \ldots, a_n$ of the $n$ nodes, constituting the graph's so-called degree 
sequence, are sampled repeatedly from the power law until $\sum_{i=1}^n a_i$ comes out even. Then $\sum_{i=1}^n a_i$ labeled balls are 
put inside an imaginary urn, where exactly $a_i$ of the balls are labeled $i$, for $1 \leq i \leq n$. A pair of balls, say of labels 
$u$ and $v$, is then withdrawn from the urn and the edge $(u,v)$ is added to the graph; this process is repeated until the urn becomes 
empty. This algorithm clearly generates a multigraph, where self-loops and multiple edges are allowed to exist. What we do as a last step is
to discard such undesirable edges, which at the end yields a random graph whose degree sequence is an approximation of the one sampled.

\begin{figure*}[p]
   \centering

   \begin{tabular}{cc}
   \includegraphics[width=\graphWidth]{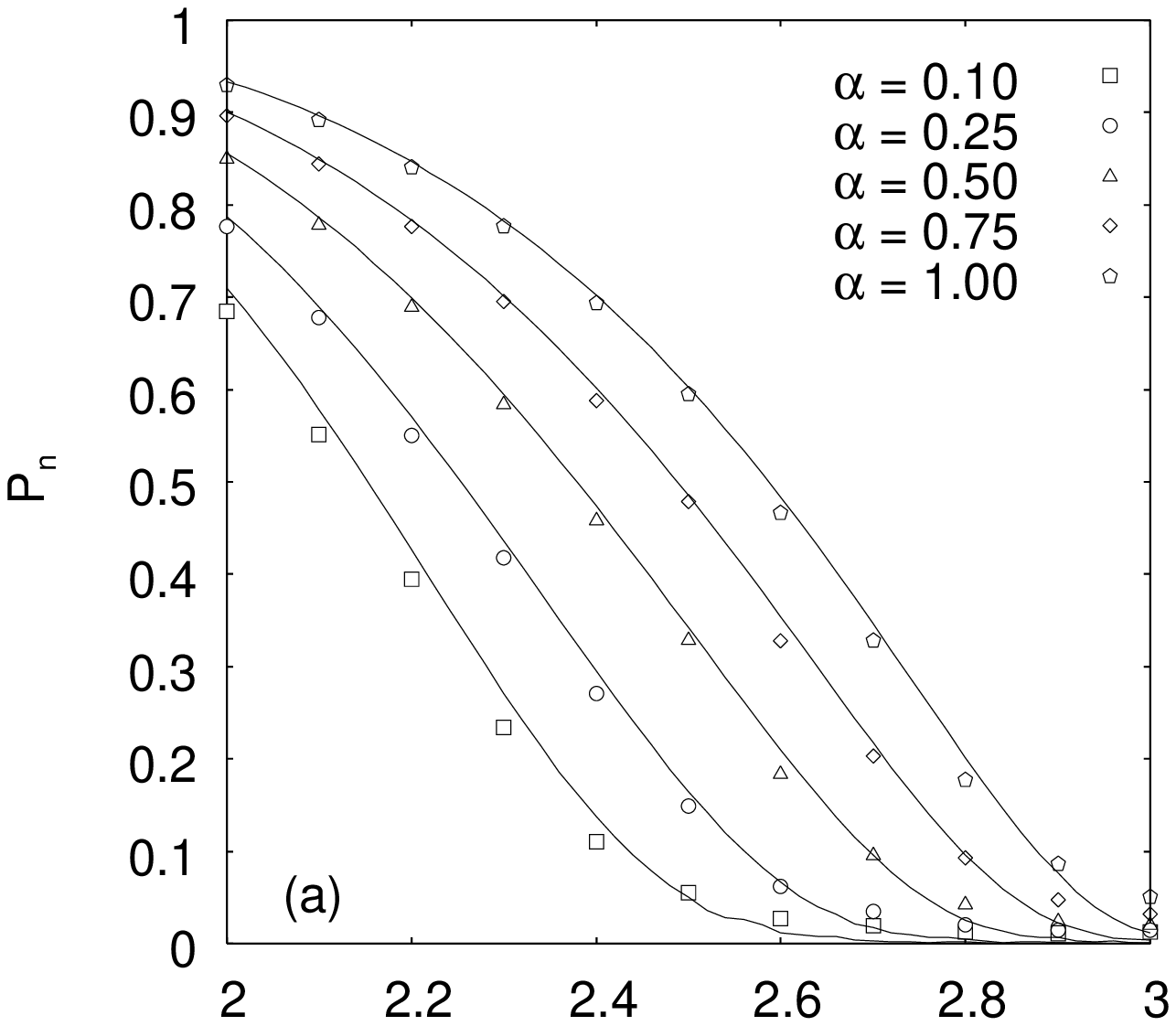}  &
   \includegraphics[width=\graphWidth]{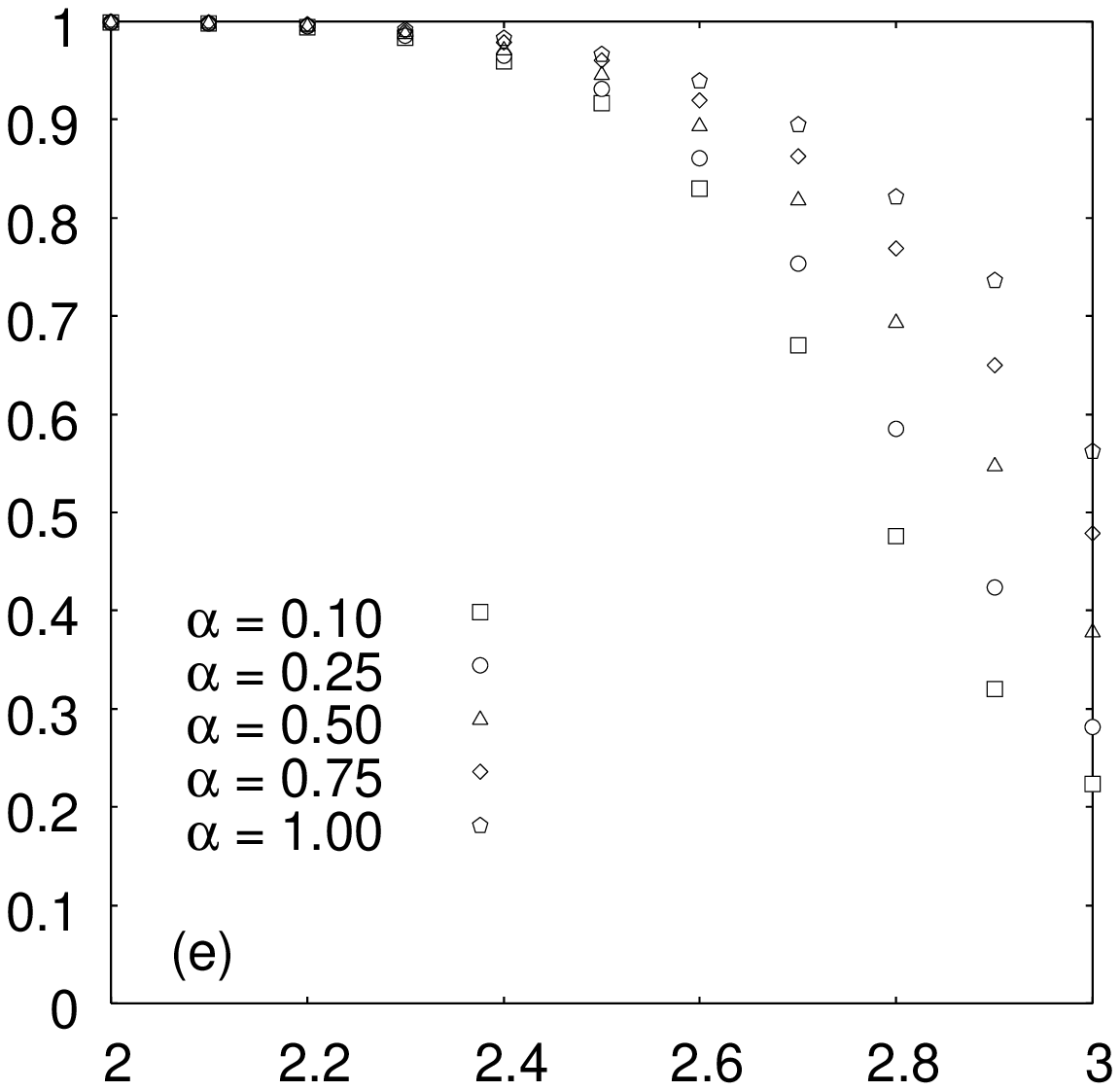}  \\
   \includegraphics[width=\graphWidth]{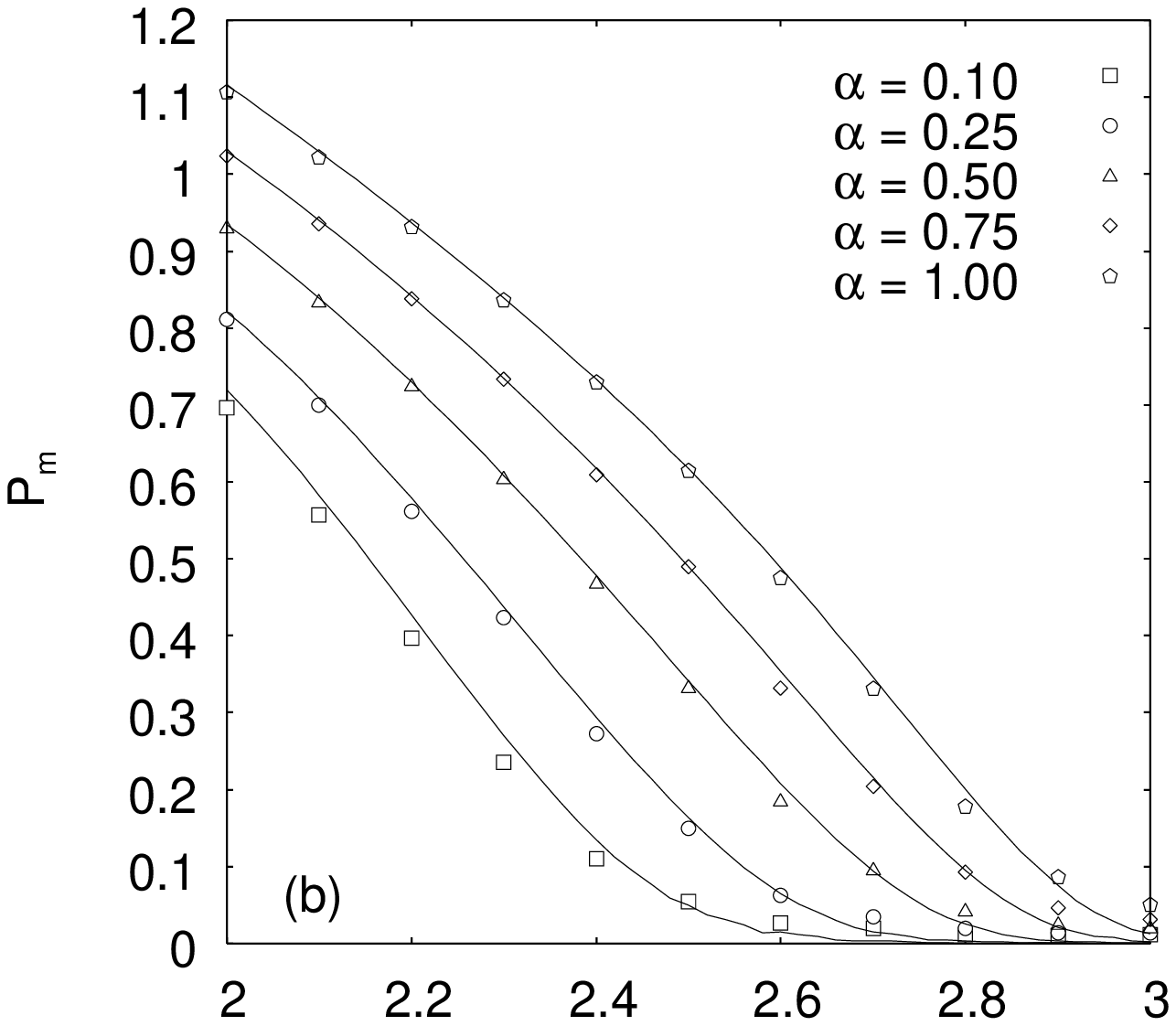}  &
   \includegraphics[width=\graphWidth]{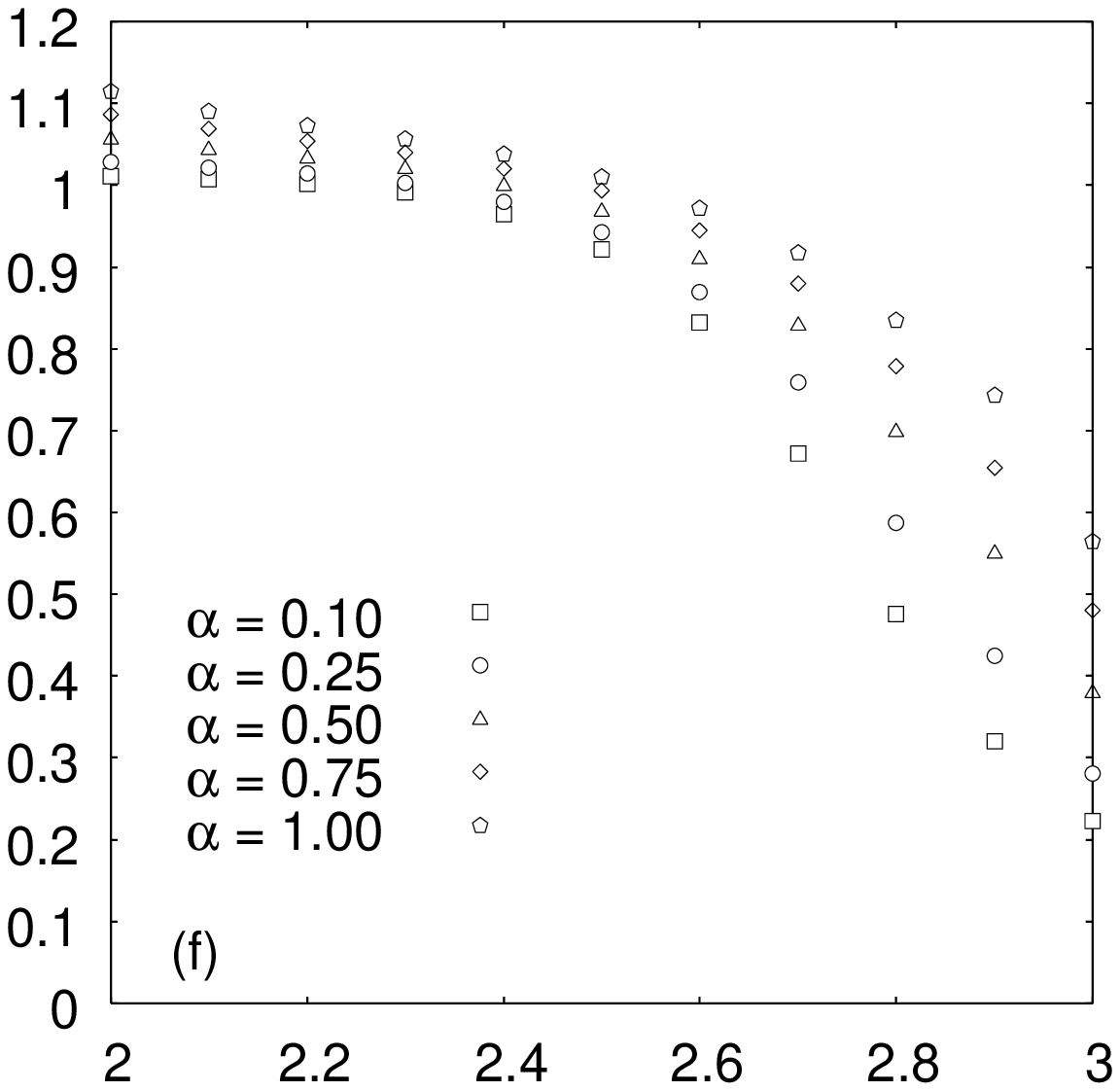}  \\
   \includegraphics[width=\graphWidth]{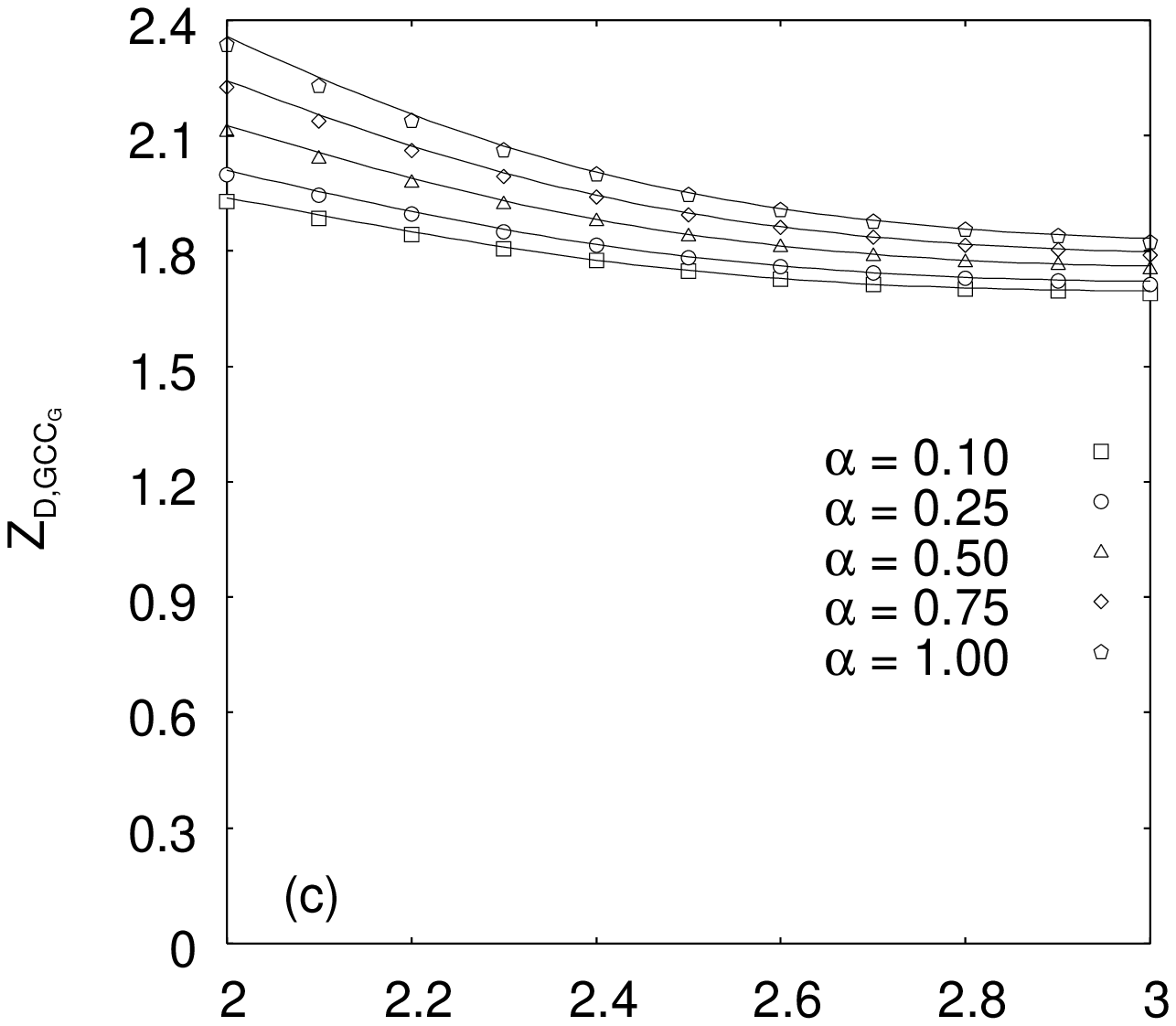}  &
   \includegraphics[width=\graphWidth]{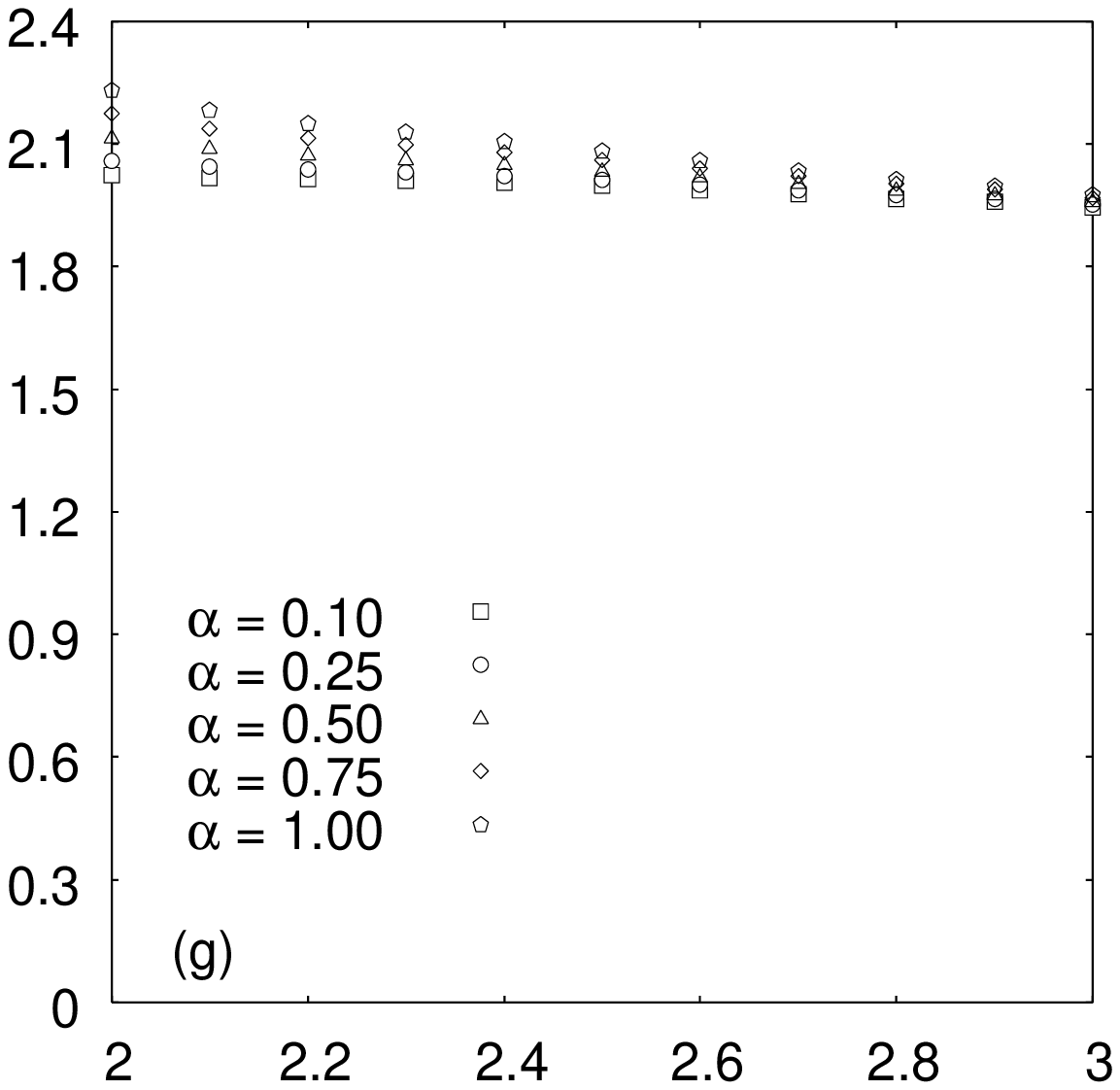}  \\
   \includegraphics[width=\graphWidth]{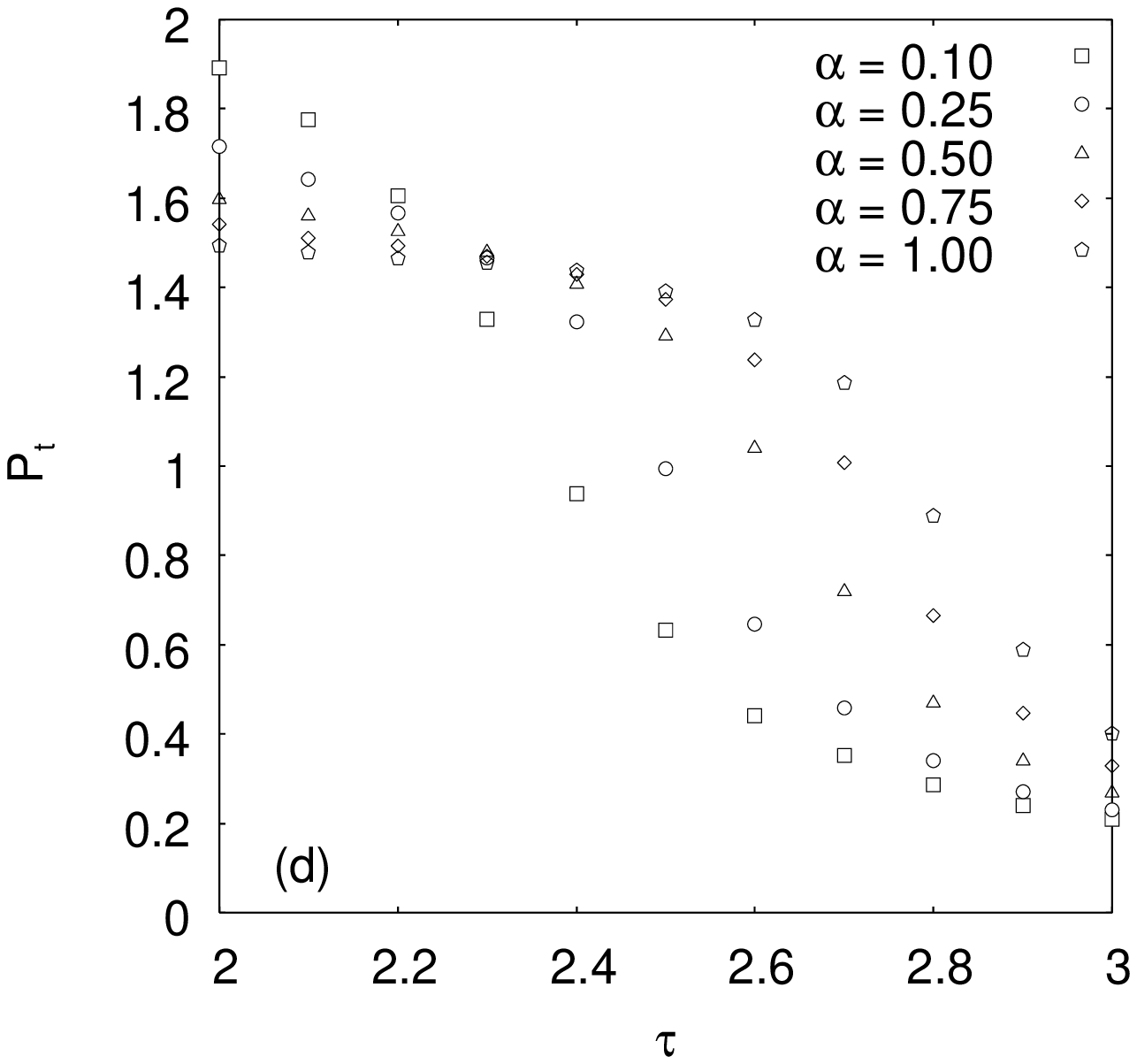} &
   \includegraphics[width=\graphWidth]{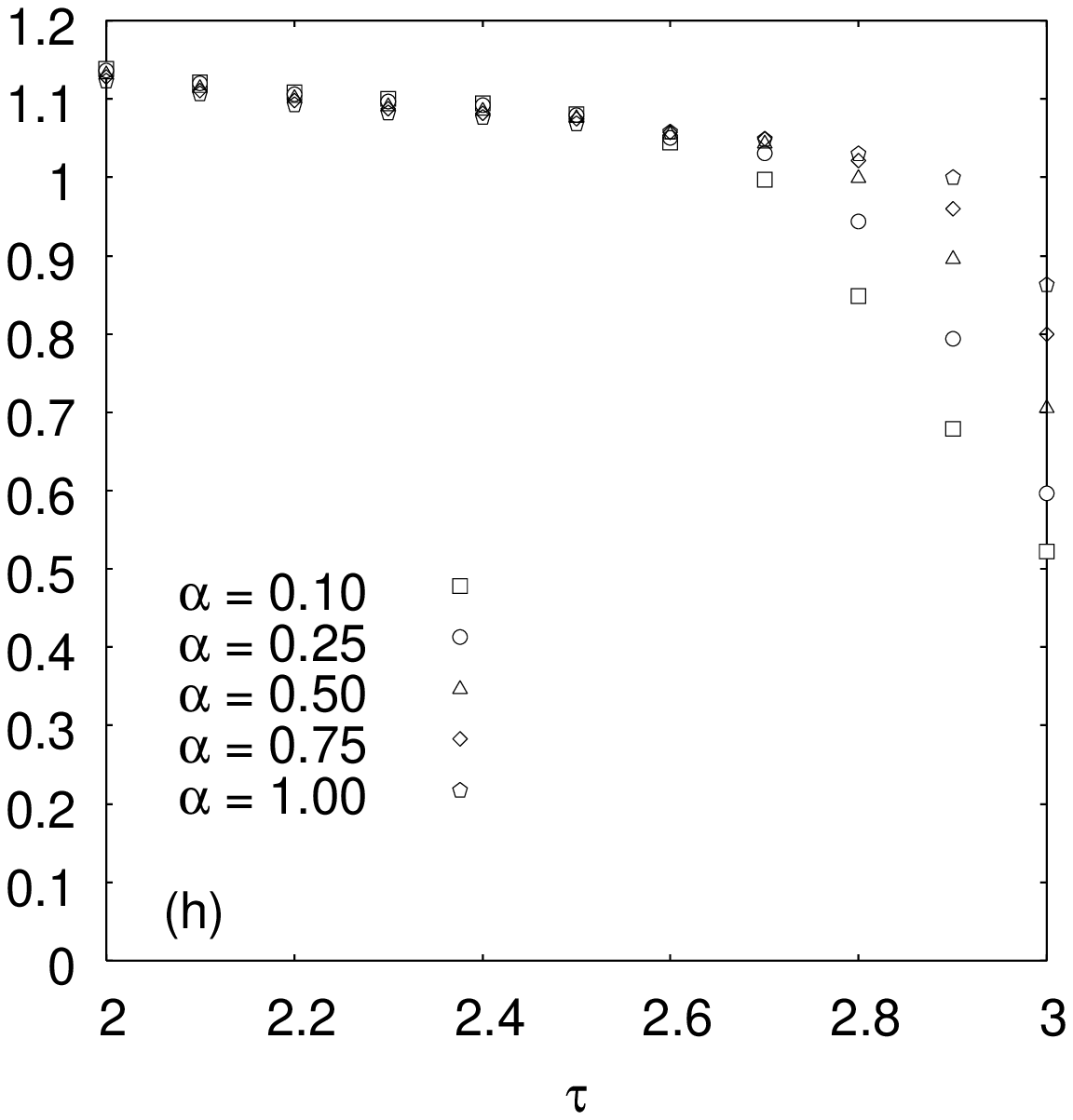}
   \end{tabular}

   \caption{Simulation results for the uniform (a--d) and the degree-based (e--h) approach on random graphs having node degrees 
   distributed as a power law. The plots show $\Pn$ (a, e), $\Pm$ (b, f), $Z_{D,\GCC{G}}$ (c, g), and $\Pt$ (d, h) for
   $\alpha=0.10,0.25,0.50,0.75,1.00$. Solid lines give the analytical predictions of Section~\ref{sec:analysis}.}
   \label{fig:sim_powerlaw}
\end{figure*}

Figure~\ref{fig:sim_powerlaw} shows simulation results for random graphs having node degrees distributed according to a power law.
The plots for the uniform approach (Figure~\ref{fig:sim_powerlaw}(a--d)) show poor results,
as $\Pn$ stays clear of $1$ for most values of $\tau$, but they do nonetheless corroborate the analytical predictions of 
Section~\ref{sec:analysis} in parts (a), (b), and (c).
For part (d) no analytical result is given, since equations (\ref{eq:zrgccg}) and (\ref{eq:zrgccd}), as similarly observed in 
\cite{newman2001}, do not converge. The plots for the degree-based approach (Figure~\ref{fig:sim_powerlaw}(e--h)), in turn, 
show excellent results for $\tau < 2.4$. In this range, $\Pn \approx 1$ and both $\Pm$ and $\Pt$ are slightly above $1$, regardless of 
the value of $\alpha$, thus demonstrating that the dissemination subgraph is very close to a spanning tree. As for $Z_{D,\GCC{G}}$,
it stays modestly valued below roughly $2.25$ throughout the entire spectrum of $\tau$ values.

\section{Resilience and adaptability} \label{sec:dynamicTopology}

\subsection{Resilience to node and link failures}

Let $\gamma_1$ and $\gamma_2$ be the probabilities, respectively, that a given node and link are operational. Letting
$\gamma = \gamma_1 \gamma_2$ be the probability that a given transmission is successful, we now consider the problem of using a
dissemination subgraph to disseminate information when each transmission has a failure probability of $1-\gamma$.
(Note, before we begin, that a simple protocol employing acknowledgement messages to ensure reliable transmissions can be used when 
$\gamma$ is substantially low. In spite of this fact, our interest is to verify what happens to the value of $\Pn$
when a failure may occur and no additional message is sent to make up for it.)

Let us consider what happens to $\GCC{G}$ when failures may occur. For such, let $G'$ be the graph obtained from
$G$ by independently removing every edge with probability $1-\gamma$. Employing the same nomenclature as in Section~\ref{sec:analysis},
a node of $G'$ is outside $\GCC{G'}$ if and only if each of its neighbors in $G$ is a dead end with respect to it in $G'$.
Considering a randomly chosen node $u$, let $q'$ be the probability that a given neighbor $v$ of $u$ in $G$ is a dead end with respect
to it in $G'$. We have
\begin{equation}
   q' = \sum_{b=1}^{n-1} \frac{b P_G(b)}{Z_G} (1 - \gamma + \gamma q')^{b-1},
\end{equation}
where $(1-\gamma +\gamma q')^{b-1}$ indicates, when $b$ is the degree of $v$, that each of the neighbors of $v$ in $G$ that are not $u$ 
either is not a neighbor of $v$ in $G'$ or is itself a dead end with respect to $v$ in $G'$. So, if $\theta_{G'}$ is the expected size of 
$\GCC{G'}$, we obtain
\begin{equation}
   \theta_{G'} = 1 - \sum_{a=0}^{n-1} P_G(a) (q')^{a}.
\end{equation}

We have carried out simulations on random graphs having $n=10000$ and degrees distributed according to either a Poisson distribution
with $1 \leq z \leq 10$ or a power law with $2 \leq \tau \leq 3$. Our aim has been to analyze $\Pn$ in the degree-based approach
when $\gamma=0.95$, i.e., when each transmission has a $0.95$ probability of success. These simulations have followed the same methodology
as in Section~\ref{sec:simulation}.

Notice that an upper bound on $\Pn$ when $\gamma > 0$ can be obtained by considering a dissemination on all the edges of $G'$. Such a bound
is thus $\theta^2_{G'}/\theta^2_{G}$. The degree-based approach to the construction of $D$ will then be as resilient to failures as
$\Pn$ is close to $\theta^2_{G'}/\theta^2_{G}$.

\begin{figure*}[!t]
   \centering

   \begin{tabular}{cc}
   \includegraphics[width=\graphWidth]{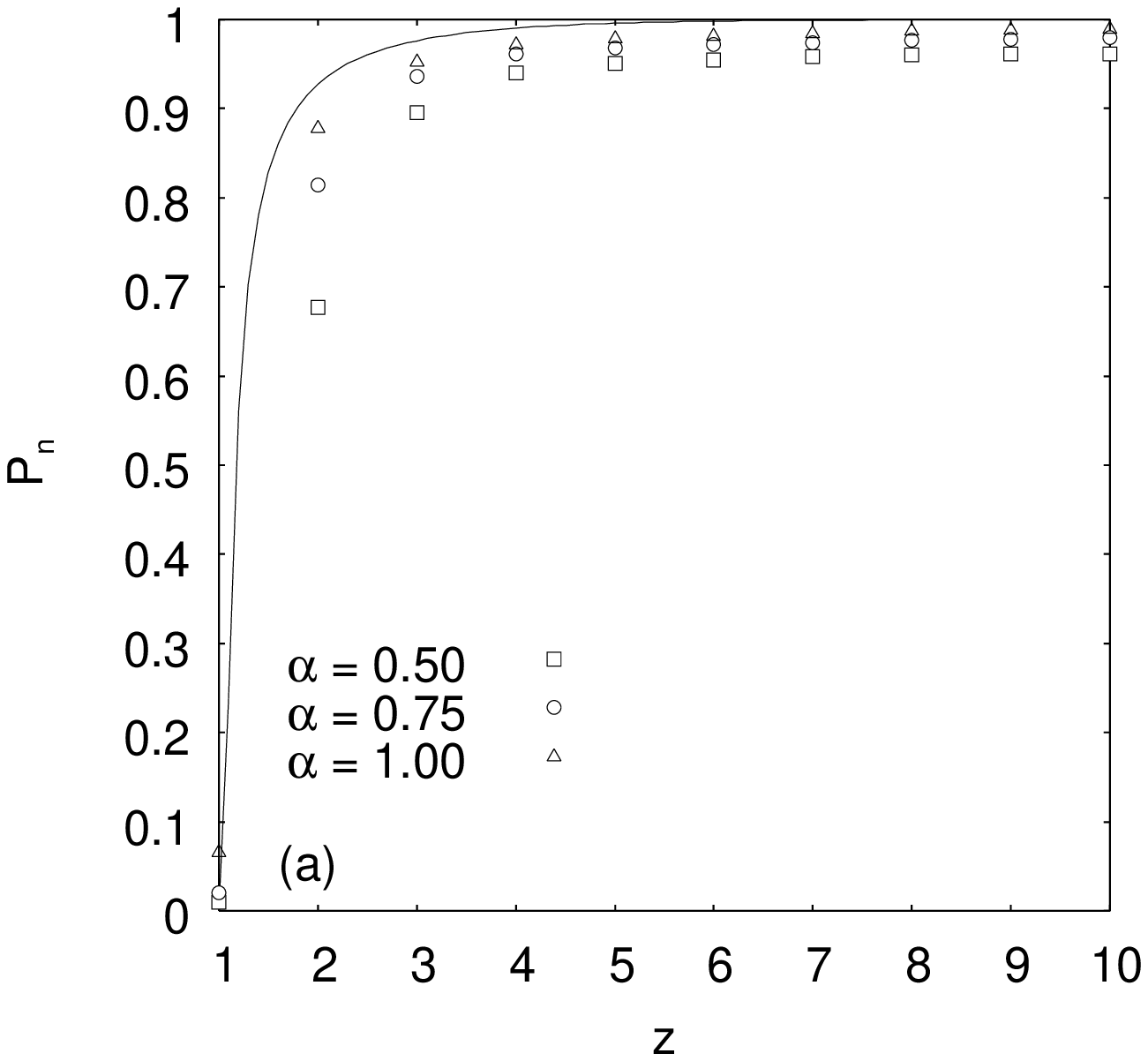}  &
   \includegraphics[width=\graphWidth]{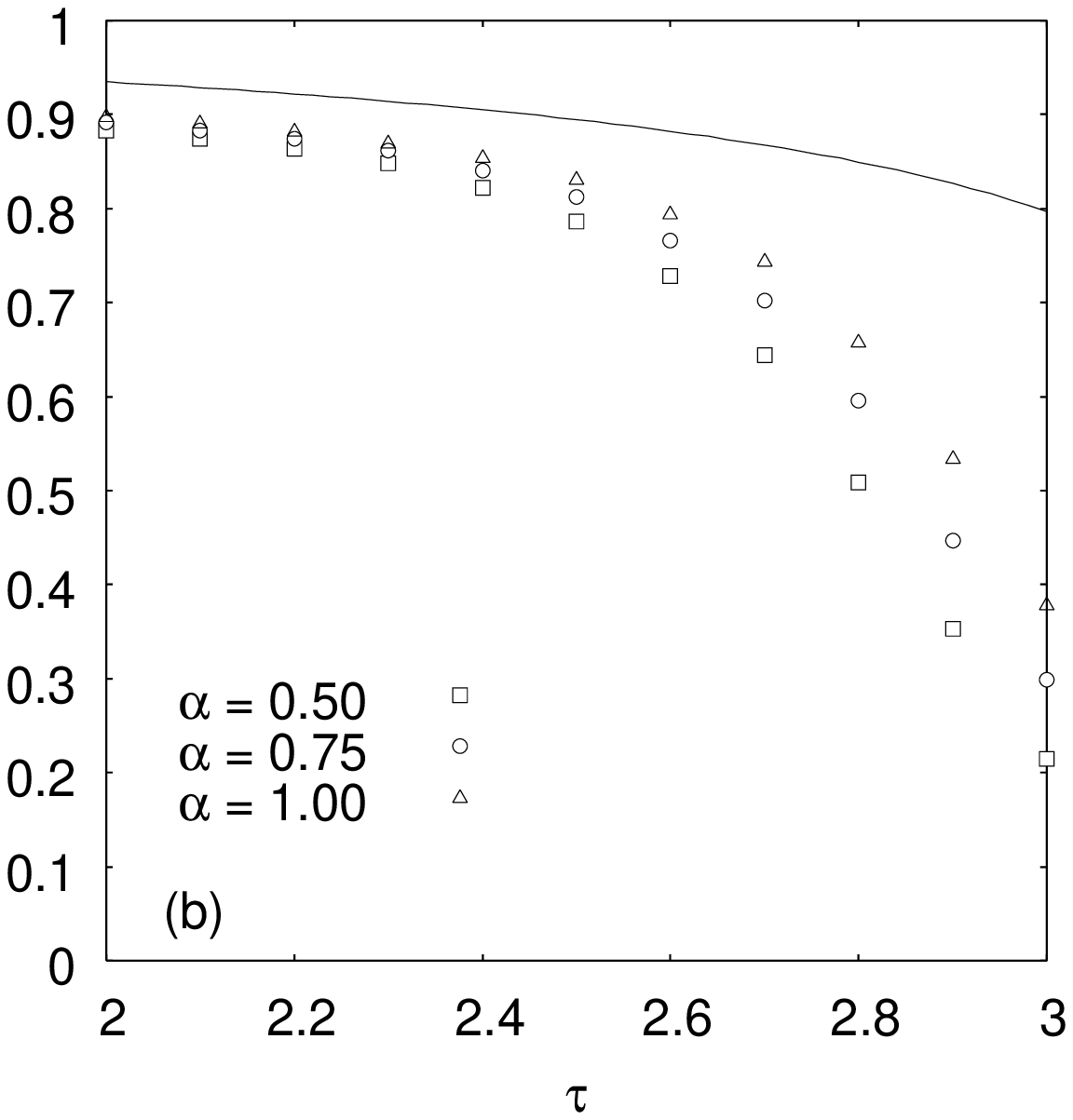}
   \end{tabular}

   \caption{Simulation results of the degree-based approach with $\gamma=0.95$ on random graphs having Poisson (a)
   and power-law (b) node-degree distributions for $\alpha=0.50,0.75,1.00$. Solid lines give the analytical predictions for 
   $\theta^2_{G'}/\theta^2_{G}$.}
   \vspace{0.2in}
   \label{fig:sim_failure}
\end{figure*}

Figure~\ref{fig:sim_failure} shows the results for $\alpha=0.50, 0.75, 1.00$ and provides an indication of how resilient the dissemination
subgraph is to transmission failures.
Clearly, in both the Poisson case (part (a) of the figure) and the power-law case (part (b)), $\Pn$ approaches $\theta^2_{G'}/\theta^2_{G}$
as $G$ gets denser (i.e., higher $z$ or lower $\tau$, as the case may be).

\subsection{Adaptability to topology changes}

We now take a brief look at how a dissemination subgraph $D$ can be made to cope with dynamic topology changes in $G$. As customary
in such cases, we model the addition or removal of a node as, respectively, the addition or removal of the edges that are incident
to it. It then suffices that we consider the addition or removal of single edges, in which context we further assume that the two
end nodes of the edge in question are capable of detecting its appearance or disappearance instantaneously.

The crux of this adaptability issue is that $D$, being constructed by strictly local actions by the nodes, can undergo changes that affect
only a small vicinity of the edge that is being added or removed (this is to be contrasted with other situations---cf., e.g., \cite{afek1995}---in
which the impact of topological changes spreads much more widely). Let $(u,v)$ be an edge that is added
to or removed from $G$. In the uniform approach, only $u$ and $v$ need remake their choices; in the degree-based approach, this holds for $u$ and $v$, 
and also for their neighbors (whose choices are affected by the degree of $u$ or $v$, as the case may be).

\section{Conclusions} \label{sec:conclusions}
In this paper we have considered the use of a spanning subgraph for disseminating a piece of information, originally known
to a single node, to all the other nodes of an unstructured network. We have introduced two local heuristics, referred to as
the uniform and the degree-based approach, for building what we call a dissemination subgraph. As we argued toward the 
end of the paper, the heuristics' intrinsically local nature leads to a degree of resilience of the dissemination subgraph
to failures, and also to a relative ease of adaptation to topological changes.

We have contributed an innovative mathematical analysis of the uniform approach, one that we hope can be extended to the degree-based 
approach as well, and also inspire the mathematical analysis of similar problems. Our simulations on random
graphs corroborate our analytical results for the uniform approach and demonstrate the efficacy, in terms of some relevant
indicators, of the degree-based approach for networks in which node degrees are distributed according to a Poisson distribution or to
a power law.

We find it remarkable that independent, strictly local decisions by the nodes of a complex network are capable of giving rise to
a global structure that in many cases comes very near a subgraph with, on average, important properties related to its
use as a substrate for information dissemination. These properties include the ability to reach nearly every node in the originator's 
connected component in the network, and
do so with relatively modest requirements concerning the overall number of messages and per-node transmission bandwidth. They also
include stretching paths only by a small factor when compared to the corresponding paths in the network.

\subsection*{Acknowledgments}

The authors acknowledge partial support from CNPq, CAPES, and a FAPERJ BBP
grant.

\bibliography{disseminationSubgraph}
\bibliographystyle{plain}

\end{document}